\documentclass[12pt]{article}

\usepackage{latexsym}

\usepackage[font=small,labelfont=bf]{caption}

\usepackage{graphics}
\usepackage{graphicx}
\usepackage{epstopdf}
\usepackage{amssymb}
\usepackage{tabularx}
\usepackage{caption}
\usepackage{subcaption}
\usepackage{hyperref}
\usepackage{tabularx}
\usepackage{gensymb}
\usepackage{mathtools,amsmath,amssymb,mathrsfs,color,esint}
\usepackage{array}
\usepackage{makecell}
\usepackage{float}
\usepackage{dcolumn}

\begin{document}


\title {Polarized image of equatorial emission in horizonless spacetimes I: traversable wormholes}

\author{
Valentin Delijski$^{1}$\footnote{E-mail: \texttt{vdelijski@phys.uni-sofia.bg}}, \,
Galin Gyulchev$^{1}$\footnote{E-mail: \texttt{gyulchev@phys.uni-sofia.bg}},  \,
Petya Nedkova$^{1}$\footnote{E-mail: \texttt{pnedkova@phys.uni-sofia.bg}},
\\ Stoytcho Yazadjiev$^{1,2}$\footnote{E-mail: \texttt{yazad@phys.uni-sofia.bg}}\\ \\
 {\footnotesize${}^{1}$ Faculty of Physics, Sofia University,}\\
  {\footnotesize    5 James Bourchier Boulevard, Sofia~1164, Bulgaria }\\
    {\footnotesize${}^{2}$ Institute of Mathematics and Informatics,}\\
{\footnotesize Bulgarian Academy of Sciences, Acad. G. Bonchev 8, } \\
  {\footnotesize  Sofia 1113, Bulgaria}}
\date{}
\maketitle

\begin{abstract}
We study the linear polarization from the accretion disk around a class of static traversable wormholes. Applying the simplified model of a magnetized fluid ring orbiting in the equatorial plane, we search for characteristic signatures, which could distinguish wormhole from black hole spacetimes by their polarization properties. For the purpose we analyse the direct polarized images for different inclination angles, the strongly lensed indirect images, and the polarized radiation which reaches the asymptotic observer through the wormhole throat, and compare to the Schwarzschild black hole. For small inclination angles the two types of compact objects lead to a very similar polarization pattern of the direct images. More significant distinctions are observed for the strongly lensed indirect images, where the polarization intensity in the wormhole spacetimes can grow up to an order of magnitude compared to the Schwarzschild black hole. Detecting radiation from the region across the wormhole throat leads to the formation of an additional structure of ring images with distinct polarization properties. The twist of the polarization vector around the ring is less pronounced, thus modifying the polarization pattern,  and the polarization intensity can increase with an order of magnitude compared to the radiation from our universe. Thus, while it could be difficult to distinguish wormhole spacetimes by their direct polarized images,  the strongly lensed images and the polarization of the radiation through the wormhole throat provide characteristic signatures which can serve as probes for horizonless objects.
\end{abstract}

\section{Introduction}

The Event Horizon Telescope opened the possibility to observe directly the supermassive black holes in the galactic centers in the radio band \cite{EHT01}-\cite{EHT6}. However, a major challenge is to interpret the radio images by answering questions about the nature of the compact  objects and the gravitational theory describing them, and as well as about the specifications of the accretion flux. These issues are hard to be determined since the  information about the gravitational field is coupled to the magneto-hydrodynamics of the accretion process in a  non-linear way depending on many parameters and physical ambiguities.

The situation is more complicated since a large variety of compact objects are characterized  by qualitatively very similar shadows. They can deviate only minimally from the Kerr black hole and the deviation frequently falls  within the limits of the experimental error.  This class of self-gravitating systems includes black holes in the modified theories of gravity \cite{Amarilla:2010}-\cite{Ahmedov:2014}, but also more exotic objects like wormholes and naked singularities \cite{Nedkova:2013}-\cite{Shaikh:2019}. In addition, it is theoretically possible that two different compact objects possess exactly the same shadow in the strict sense, proving that the shadow boundary cannot determine the compact object uniquely \cite{Herdeiro:2021}.

In order to constrain further the spacetime geometry, we should consider additional properties of the images. We can analyse, for example,  the apparent structure of the accretion disk and the observable intensity of its emission. However, the investigation of the relativistic images of some simplified models of accretion disks demonstrates that the problem still contains a lot of degeneracy. The accretion disks around compact objects of very different physical nature closely resemble each other and we obtain again a large sample of Kerr black hole mimickers \cite{Nedkova:2019}-\cite{Joshi:2022}.

Another approach to break the degeneracy and differentiate more effectively between different compact objects by means of their radio images is to consider the polarization of the electromagnetic radiation from the accretion disk. The polarization characterizes the structure of the magnetic field in the strong-gravity region probing in particular the interaction of the local magnetic field with the spacetime geometry. In this way it brings supplementary information in addition to the observable intensity of the accretion disk.

Recently, the Event Horizon Telescope Collaboration published a survey of the observable polarization from the radio source M87*, revealing a linearly polarized emission on event-horizon scales, which is most probably produced by synchrotron radiation \cite{EHT7}-\cite{EHT8}. The observed images are interpreted by comparing with the polarization structure which would follow theoretically from different scenarios of the accretion disk physics by means of performing general relativistic magnetohydrodynamic (GRMHD) simulations. The analysis strongly suggests that the magnetically arrested disk models (MAD), which are characterized by dynamically important magnetic fields in the emission region, are most relevant for describing the vicinity of M87*.

In addition, a simple analytical toy model was developed and compared with the observational data \cite{Narayan:2021}. The polarized emission is represented as the synchrotron radiation, which originates from a thin ring of gas while orbiting in the gravitational field of the Schwarzschild black hole assuming that the local magnetic field is constant. The superposition of the radiation produced by several copies of such fluid rings builds up the emission from the innermost region of the accretion disk.  The model is characterized only by a few parameters which include the radius of the ring and its tilt with respect to the observer, the velocity of the fluid in the local rest frame and the magnitude and the direction of the magnetic field. However, despite its simplicity it leads to similar predictions as the fully non-linear MAD models, and manages to reproduce the basic features of the observable polarization.

The model of a synchrotron emitting ring was generalized to the Kerr black hole in the context of the EHT observations \cite{Gelles:2021} and extended to the modified theories of gravity \cite{Qin:2021}. The polarization structure for rotating black holes with near-extremal angular momenta was considered recently in \cite{Lupsasca:2020}, while \cite{Himwich:2020}  investigated the universal properties of the strongly lensed  polarimetric images in the vicinity of the photon ring. We should note, however, that semi-analytical studies of the linear polarization  and the lensed images of an equatorial circular emitter in black hole spacetimes date back already from the 70s and have resulted in a range of classical works \cite{Bardeen:1972}-\cite{Agol:1997} .

Our aim in this paper is to explore the polarization signatures in the spacetime of horizonless compact objects. On the one hand, we search for qualitatively new features of the polarized images which could distinguish exotic compact objects observationally. On the other hand, we probe how sensitive is the observable polarization to the spacetime geometry and how effectively we can use its structure in order to determine the physical nature of the compact objects at the galactic centres.
For the purpose we apply the analytical model of a synchrotron emitting ring of gas in the gravitational field of a  class of static traversable wormholes and investigate whether the non-black hole nature of spacetime would leave imprints on the properties of the polarization.  Since we would like to isolate effects which can be attributed primarily to the absence of an event horizon, we select geometries possessing  similar structure of the circular geodesics as the Schwarzschild black hole. In particular, the wormhole solutions contain a single photon ring, which is located at a similar radial distance as for the Schwarzschild black hole. Such relation applies also for the marginally bound and the marginally stable particle orbits, which determine the characteristic distances for the accretion process.

The paper is organized as follows. In the next section we present the static traversable wormholes which we consider and some of the characteristics of the geodesic motion, which are relevant for our studies. In section 3 we describe the physical model of the linear polarization, which results from synchrotron radiation propagating in curved spacetime and the computational procedure for obtaining its observable image. We show that in static spherically symmetric spacetime the procedure is completely algebraic due to the presence of Killing and Killing-Yano tensors.  In section 4 we present our simulated images for the linear polarization in wormhole geometry and discuss their properties in comparison to the Schwarzschild black hole. We consider direct images at different inclination angles, as well as the strongly lensed indirect images.  In the last section we summarize our results.



\section{Static traversable wormholes}

In their seminal paper \cite{Morris} Morris and Thorne suggested the general geometry which describes a static traversable wormhole

\begin{equation}\label{metric}
    ds^2 = -N^2dt^2 +\left(1-\frac{b}{r}\right)^{-1} dr^2 +r^2\left(d\theta^2+ \sin^2\theta d\phi^2\right),
   \end{equation}
The metric functions depend only on the radial coordinate $r$, and we assume that the redshift function $N(r)$ is finite and non-zero in all the coordinate range. Then, the metric describes a completely regular spacetime and we  can infer the properties of this geometry by considering the embedding of the constant $t$ and $\theta$  cross-sections in the three-dimensional Euclidean space. The embedding  describes a tunnel connecting two regions in the spacetime with a throat located at the constant radius $b(r)=r_0$ where the metric function $g_{rr}$ becomes divergent. It  further possesses the characteristic flaring out shape of the wormhole if we require that the condition $db/dr<1$ is satisfied at the throat. In addition, the two ends are asymptotically flat provided that the metric functions behave as
\begin{eqnarray}
N &=& 1-\frac{M}{r}+\mathcal O\left(\frac{1}{r^2}\right),\quad~~~\frac{b}{r} =\mathcal O\left(\frac{1}{r}\right),
\end{eqnarray}
at the spacetime infinity, where $M$ is the ADM mass of the solution.

In this work we consider a particular class of wormhole solutions with metric functions given by
\begin{eqnarray}\label{wormhole}
 N = \exp\left(-\frac{r_0}{r} - \alpha\frac{r^2_0}{r^2}\right),  \quad~~~b=r_0,
 \end{eqnarray}
where $r_0$ and $\alpha$ are positive constants. The parameter $r_0$ determines the ADM mass of the wormhole, while the parameter $\alpha$ controls the redshift of the solution. For convenience we can further represent the metric in a dimensionless form by making the conformal transformation and rescaling by the wormhole mass
\begin{eqnarray}
dS^{2}= r_0^{-2}ds^{2}, \quad~~~t\rightarrow r_0t, \quad~~~r\rightarrow r_0r. \nonumber
\end{eqnarray}
In this way we obtain a wormhole solution with a unit mass and a throat located at $r=1$.

The described spacetime possesses a photon sphere located at the maximum of the effective potential for the null geodesics

\begin{equation}
V^{ph}_{eff}= L^2\frac{N(r)}{r^2},
\end{equation}
where $L$ is the specific angular momentum of the photon, i.e. satisfying  $V^{ph}_{eff}=0$,  $dV^{ph}_{eff}/dr=0$ and $dV^{ph}_{eff}/dr<0$. We can further obtain the position of the marginally stable timelike orbit, corresponding to the inflection point of the effective potential  for the timelike geodesics

\begin{equation}
V_{eff}= N(r)\left(1 +\frac{L^2}{r^2}\right),
\end{equation}
as a solution to the equations $V_{eff}=0$, $dV_{eff}/dr=0$ and $d^2V_{eff}/dr^2=0$. The locations of the photon sphere and the marginable stable orbit can be given explicitly in the form

\begin{eqnarray}
&&r_{\text{ph}} = \frac{M}{2}\left(1 + \sqrt{1+8\alpha}\right),  \nonumber \\[3mm]
&&r_{\text{ms}} = 2M\left[\sqrt{\frac{4}{9}\left(6\alpha + 1\right)}\cosh{\left(\frac{1}{3}\operatorname{arcosh}{\frac{1+9\alpha + \frac{27}{2}\alpha^2}{(6\alpha + 1)^{\frac{3}{2}}}}\right)} + \frac{1}{3}\right].
\end{eqnarray}

On the other hand, the marginally bound timelike orbit is determined by the condition

\begin{eqnarray}
E =\frac{N^2(r)}{\sqrt{N^2(r) - rN(r)N^{'}(r)}} =1,
\end{eqnarray}
where $E$ is the specific energy of the orbiting particle.

\begin{figure*}[t!]
    \centering
    \begin{subfigure}[t]{0.6\textwidth}
        \includegraphics[width=\textwidth]{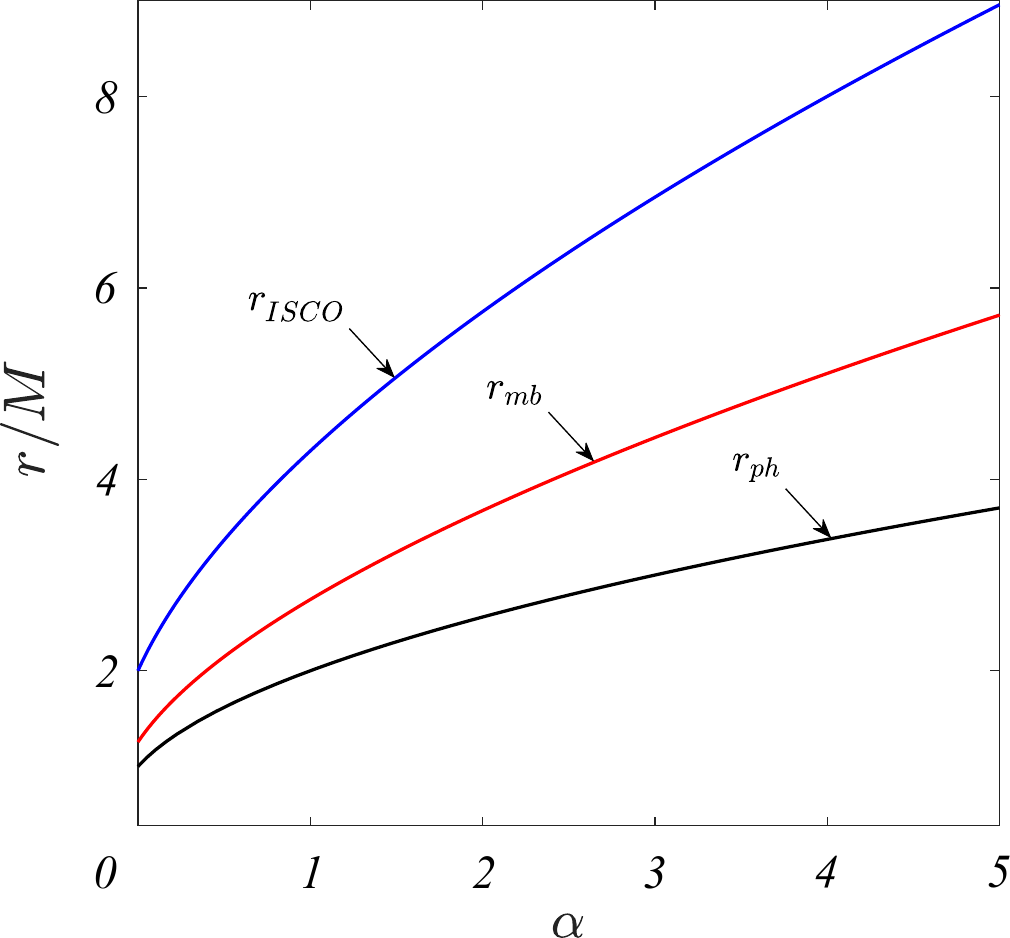}
           \end{subfigure}
           \caption{\small Location of the photon sphere $r_{ph}$, the marginally bound orbit $r_{mb}$ and marginally stable orbit $r_{ms}$ for traversable wormholes with different redshift parameter $\alpha$. }\label{fig:geodesics}
\end{figure*}

The behavior of the characteristic circular orbits for the wormhole solution ($\ref{wormhole}$) is summarized in Fig. $\ref{fig:geodesics}$. We observe that in the range $\alpha\in [2,3]$ their locations approach the corresponding values for the Schwarzschild solution. For example, for $\alpha = 2.2$ the marginally stable orbit is located at $r=6M$, for $\alpha = 2.4$ the marginally bound orbit corresponds to $r=4M$, while for $\alpha = 3$ we obtain the photon sphere at the radial distance $r=3M$.

\section{Polarized image of an equatorial emitting ring in static spherically symmetric spacetime}

We consider a ring of magnetized fluid orbiting in the equatorial plane around a static spherically symmetric compact object. Due to the interaction with the local magnetic field the fluid emits linearly polarized synchrotron radiation. In this section we will discuss the apparent polarization of the electromagnetic radiation generated in this setting as seen by an asymptotic observer and describe the general procedure for obtaining its observable  pattern and intensity.

Let us consider a static spherically symmetric metric in the general form

\begin{eqnarray}
ds^2 = - e^{2\nu(r)}dt^2 + e^{2\lambda(r)}dr^2 + R^2(r) (d\theta^2 + \sin^2\theta d\phi^2).
\end{eqnarray}

For our purpose it is convenient to express the polarization in the fluid rest frame by defining a local orthonormal tetrad  $\{e^{\mu}_{(a)}\}$ aligned with the rotating fluid element at a given radius $r$ and azimuthal angle $\phi$. The natural tetrad adapted to the spacetime symmetries is given by

\begin{eqnarray}\label{eq:tetrad}
     e_{(t)} =  e^{-\nu(r)}\partial_t, \,\,\,\,\,
    e_{(r)} = e^{-\lambda(r)}\partial_r, \,\,\,\,\,
    e_{(\theta)} =  \frac{1}{R(r)} \partial_{\theta}, \,\,\,\,\,
    e_{(\phi)} = \frac{1}{R(r)\sin\theta}\partial_{\phi}.
\end{eqnarray}

We further allow for the fluid to move with velocity $\vec{\beta}$ in the $(r)$-$(\phi)$ plane of the local frame, which we parameterize as

\begin{equation} \label{eq:betaboost}
\vec{\beta} = \beta\left(\cos\chi\,(r)+\sin\chi\,(\phi)\right).
\end{equation}
Defining a local rest frame attached to the boosted emitter

\begin{eqnarray}
\hat{e}^{\,\mu}_{(a)} = \Lambda^{\hspace{0.3cm}(b)}_{(a)}e^{\mu}_{(b)},
\end{eqnarray}
vectors are transformed by the  Lorenz transformation $\Lambda^{(a)}_{\hspace{0.3cm}(b)}$
\begin{equation} \label{eq:boostedframe}
\hat{V}^{(a)} = \Lambda^{(a)}_{\ \ (b)} V^{(b)},
\end{equation}
where  the matrix $\Lambda^{(a)}_{\hspace{0.3cm}(b)}$ is given in explicit form as
\begin{eqnarray}
    \small
    &&\Lambda
        =\begin{pmatrix}
             \gamma  & -\beta  \gamma  \cos \chi & -\beta  \gamma  \sin \chi & 0 \\
             -\beta  \gamma  \cos \chi & (\gamma -1) \cos ^2\chi+1 & (\gamma -1) \sin \chi \cos \chi & 0 \\
             -\beta  \gamma  \sin \chi & (\gamma -1) \sin \chi \cos \chi & (\gamma -1) \sin ^2\chi+1 & 0 \\
             0 & 0 & 0 & 1 \\
        \end{pmatrix}, \\[2mm]
    \normalsize
    &&\gamma =(1-\beta^2)^{-1/2}
\end{eqnarray}
Given a vector $\hat{V}^{(a)}$ in the local rest frame, we can recover its components in the spacetime coordinates  $V^{\mu}$ by the inverse transformation

\begin{equation} \label{vec_inverse}
V^{\mu} = \hat{e}^{\,\mu}_{(a)} \hat{V}^{(a)}.
\end{equation}

In the local rest frame  we introduce the magnetic field  $\vec{B} =\left(\hat{B}^{ r}, \hat{B}^{\phi}, \hat{B}^{\theta}\right)$  and denote the local three momentum by $\vec{p} =\left(\hat{p}^{ r}, \hat{p}^{\phi}, \hat{p}^{\theta}\right)$. Under the action of the magnetic field the fluid emits synchrotron radiation with polarization

\begin{equation} \label{eq:polcross}
  \vec{f} = \frac{\vec{p} \times \vec{B}}{|\vec{p}|},
\end{equation}
where we have $\hat{f}^{(t)} =0$. For convenience we further normalize the polarization vector to satisfy

\begin{equation}
\hat{f}^{a}\hat{f}_{a} = \sin^2\zeta |\vec{B}|^2,
\end{equation}
denoting the angle between $\vec{p}$ and the magnetic field $\vec{B}$ by $\zeta$

\begin{equation} \label{eq:sinzeta}
\sin \zeta = \frac{| \vec{p} \times \vec{B}|}{|\vec{p}||\vec{B}|}.
\end{equation}

The components of the polarization 4-vector $f^\mu$ in the spacetime coordinates are recovered in the standard way by performing the transformation ($\ref{vec_inverse}$).

Having defined the polarization at the point of emission we should calculate how it transforms while the electromagnetic radiation propagates through the curved spacetime in order to reach the asymptotic observer. According to the geometric optics approximation the right rays propagate on null geodesics, while the polarization $4$-vector is parallel transported along the geodesic, and it is normal to the $4$-momentum $p^\mu$, i.e.

\begin{eqnarray}
&&p^\mu \nabla_{\mu} p_\nu =0 \nonumber \\[2mm]
&&p^{\mu}\nabla_{\mu}f_\nu=0, \;\;\; p^\mu f_\mu=0.
\end{eqnarray}

In general, in order to obtain the polarization vector  at the observer's position we have to solve the described differential equations with the corresponding boundary conditions. However, in static spherically symmetric spacetime the procedure reduces to algebraic manipulations due to the high degree of symmetry. The spacetime possesses four Killing vectors which make the geodesic motion completely integrable. In addition, we have the hidden symmetries in the form of irreducable Killing and Killing-Yano tensors which don't generate isometries but lead to conserved constants of motion. Using their explicit form we can derive an algebraic expression for  the polarization $4$-vector at any point of the spacetime.

The static spherically symmetric spacetime admits a second order Killing-Yano tensor $Y_{\mu\nu}$ with the following non-zero components in the standard coordinate basis

\begin{eqnarray}
Y_{\theta\phi}=-Y_{\phi\theta}=R^3(r) \sin\theta.
\end{eqnarray}
	
In addition we have a conformal Killing-Yano tensor  ${\tilde Y}_{\mu\nu}$ arizing as the Hodge dual of $Y_{\mu\nu}$, i.e. $ {\tilde Y}_{\mu\nu} = \frac{1}{2}\epsilon_{\mu\nu\alpha\beta}Y^{\alpha\beta}$. Using $Y_{\mu\nu}$ and ${\tilde Y}_{\mu\nu}$  we can construct two constants of motion $\kappa_1$ and $\kappa_2$ which are conserved along the photon trajectory and related to the photon's 4-momentum $p^{\mu}$ and polarization vector $f^{\mu}$ in the following way

\begin{eqnarray}
\kappa_1= \frac{1}{2} {\tilde Y}_{\mu\nu}p^\mu f^\nu , \;\;\;    \kappa_2= Y_{\mu\nu} p^\mu f^\nu  .
\end{eqnarray}

In explicit form we have

\begin{eqnarray}\label{kappa_12}
&&\kappa_1= R(r) e^{\nu(r) +\lambda(r)} (p^t f^r - p^rf^t), \nonumber  \\[2mm]
&&\kappa_2=  R^3(r) \sin \theta (p^\theta f^\phi - p^\phi f^\theta).
\end{eqnarray}

These expressions allow us to determine the observable polarization entirely in terms of the boundary data on the geodesic. Let us consider a light ray emitted at spacetime coordinates $x^\mu_s$ with linear polarization $f^\mu_s$ which reaches a distant observer located at  $x^\mu_{obs}$ with polarization  $f^\mu_{obs}$. In order to obtain the observable polarization we should define a local reference frame at the observer's position. We introduce the natural orthonornal tetrad given by ($\ref{eq:tetrad}$) and denote by $p^{(a)}$ the projection of the $4$-momentum in the local basis. The projection is completely determined by two angles $\alpha$ and $\beta$ defined as \cite{Bardeen}

\begin{eqnarray}
\sin\alpha = \frac{p^{(\theta)}}{p^{(t)}}, \quad \,\,\, \tan\beta = \frac{p^{(\phi)}}{p^{(r)}},
\end{eqnarray}
which we can choose as celestial coordinates in the observer's sky. Since $\alpha$ and $\beta$ scale inversely with the radial coordinate at large distances, it is convenient to rescale them defining the celestial coordinates $x =  r \alpha$ and $y=  r\beta$. For an asymptotic observer we can take the limit $r\rightarrow\infty$ and the celestial coordinates reduce to

\begin{eqnarray}\label{obs_coord}
x &=& -R(r)p^{(\phi)} = -\frac{p_{\phi}}{\sin\theta_0}, \nonumber \\[2mm]
y &=&  R(r)p^{(\theta)} = p_{\theta},
\end{eqnarray}
where $\theta_0$ is the inclination angle of the observer.

The constants of motion $\kappa_1$ and $\kappa_2$ can be expressed by means of the projection of the $4$-momentum and the polarization vector in the local reference frame of the observer

\begin{eqnarray}\label{CONPOL}
&&\kappa_1=R(r) (p^{(t)}f^{(r)} - p^{(r)}f^{(t)} ), \nonumber  \\[2mm]
&&\kappa_2=R(r)(p^{(\theta)}f^{(\phi)} - p^{(\phi)}f^{(\theta)}).
\end{eqnarray}

Using the orthogonality condition $p^\mu f_\mu=0$ we can solve (\ref{CONPOL}) for $f^{(\theta)}$ and $f^{(\phi)}$ and obtain

\begin{eqnarray}
f^{(\theta)} = \frac{p^{(\theta)}}{(p^{(\theta)})^2 + (p^{(\phi)})^2}\left[- \frac{\kappa_1}{R} \frac{p^{(r)}}{p^{(t)}} + \left((p^{(t)})^2 - (p^{(r)})^2   \right) f^{(t)}    \right] - \frac{p^{(\phi)}}{(p^{(\theta)})^2 + (p^{(\phi)})^2} \frac{\kappa_2}{R}, \\[3mm] \nonumber
f^{(\phi)} = \frac{p^{(\phi)}}{(p^{(\theta)})^2 + (p^{(\phi)})^2}\left[- \frac{\kappa_1}{R} \frac{p^{(r)}}{p^{(t)}} + \left((p^{(t)})^2 - (p^{(r)})^2   \right) f^{(t)}    \right] + \frac{p^{(\theta)}}{(p^{(\theta)})^2 + (p^{(\phi)})^2} \frac{\kappa_2}{R}.
\end{eqnarray}

For an asymptotic observer we can simplify these relations taking advantage of a further constant of motion associated with the hidden symmetries in static spherically symmetric spacetimes. The square of the Killing-Yano tensor $Y_{\mu\nu}$ gives rise to the Killing tensor $K_{\mu\nu}=Y_{\mu\alpha} Y_{\nu}^{\;\;\alpha}$, which generates the conserve quantity $C=K_{\mu\nu}p^\mu p^\nu$ along the photon trajectories. In explicit form we have

\begin{eqnarray}
C=K_{\mu\nu}p^\mu p^\nu  = R^4(r)\left[(p^\theta)^2 + \sin^2\theta (p^\phi)^2\right] = R^2(r)\left[(p^{(\theta)})^2 + (p^{(\phi)})^2\right],
\end{eqnarray}

Using this integral of motion and the property of the null geodesics $p_\mu p^\mu=0$ we can show that in the limit $r\rightarrow\infty$ the terms containing the polarization component $f^{(t)}$ vanish.  Then, we take into account the definition of the celestial coordinates ($\ref{obs_coord}$)
and obtain the following expressions for the components of the polarization vector in the coordinate system on the asymptotic observer's screen

\begin{eqnarray}
f^x = \frac{x\kappa_1  + y\kappa_2}{x^2 + y^2} , \nonumber \\[2mm]
f^y = \frac{y\kappa_1 - x\kappa_2}{x^2 + y^2} .
\end{eqnarray}.
We can further normalize the polarization vector measured by the observer to unity as

\begin{eqnarray}\label{obs_polarization}
f^x_{obs} = \frac{x\kappa_1  + y\kappa_2}{\sqrt{(\kappa_1^2 + \kappa_2^2)(x^2 + y^2)}} , \nonumber \\[2mm]
f^y_{obs} = \frac{y\kappa_1  - x\kappa_2}{\sqrt{(\kappa_1^2 + \kappa_2^2)(x^2 + y^2)}} .
\end{eqnarray}

In this way we obtain expressions for the observable polarization at infinity which depend only on the constants of
motion $\kappa_1$ and $\kappa_2$ and the projection of the photon trajectory on the observer screen $(x, y)$. The conserved quantities $\kappa_1$ and $\kappa_2$ can be determined by the initial data for the polarization, i.e. by the parameters of the physical process which gives rise to it in the accreting fluid. We consider the components of the polarization vector in the fluid local rest frame $\hat{f}^{(a)}$ following from the synchrotron radiation ($\ref{eq:polcross}$). Then, we project them in the local tetrad attached to the orbiting fluid at certain radius $r_s$ ($\ref{eq:tetrad}$) and substituting in ($\ref{kappa_12}$) we obtain expressions for $\kappa_1$ and $\kappa_2$ only in terms of the initial data in the fluid rest frame

\begin{eqnarray}\label{kappa_rfr}
\kappa_1= \gamma R(r_s) \left[{\hat p^{(t)}}{\hat f^{(r)}}  + \beta \left({\hat p^{(r)}}{\hat f^{(\theta)}} - {\hat f^{(r)}}{\hat p^{(\theta)}} \right) \right], \nonumber \\[2mm]
\kappa_2 = \gamma R(r_s) \left[\left({\hat p^{(\theta)}}{\hat f^{(\phi)}} - {\hat f^{(\theta)}}{\hat p^{(\phi)}} \right)  - \beta {\hat p^{(t)}}{\hat f^{(\phi)}}\right].
\end{eqnarray}
As we defined previously in our model, $\beta$ is the physical velocity of the orbiting emitter and $\gamma = 1/\sqrt{1-\beta^2}$.

In order to examine the polarization structure we use two observables - the polarization intensity and the direction of the polarization vector. The polarization intensity is proportional to the norm of the observable polarization vector but we need to take into account additional factors determined by the phenomenology of the synchrotron emission \cite{Narayan:2021}. The intensity of the synchrotron radiation depends on the angle $\sin\zeta$ between the direction of the magnetic field and the 4-velocity in the fluid rest frame, which is modulated by the frequency of the emitted photons $\alpha_\nu$ to a power law $(\sin\zeta)^{1+\alpha_\nu}$. Since this is the emission per unit volume, it should be further multiplied by the geodesic path $l_p$ length in the emitting region. For an optically and geometrically thin disk, which we assume in our model, the geodesic path length
can be expressed as

\begin{eqnarray}
 l_p=\frac{\hat p^{(t)}}{\hat p^{(\theta)}}\,H,
\end{eqnarray}
by means of the 4-momentum in the fluid rest frame and the disk height $H$. While propagating in the spacetime in order to reach the observer,  the intensity is further Doppler-boosted by a factor $\delta^{3+\alpha}$, where $g$ is the gravitational redshift

\begin{equation}
  g = \frac{E_{obs}}{E_s} = \frac{1}{\hat p^{(t)}},
\end{equation}
given by ratio of the photon's energy measured at the observer's frame $E_{obs}$ and the emitter's frame $E_s$. Thus, assuming that the polarization vector in the observer's frame is normalized to unity,  the observed linear polarization intensity is given by the expression
\begin{eqnarray}
|P| = \delta^{3+\alpha}\, l_p \, (\sin\zeta)^{1+\alpha_\nu},
\end{eqnarray}
up to a proportionality constant. Models of M87* are consistent with the power $\alpha_\nu=1$, therefore in our studies we adopt this value.

Since the intensity is proportional to the norm of the polarization vector we can define the components of the observed electric field as

\begin{eqnarray}
E^x_{obs} &=& \delta^2\, l_p^{1/2}\, \sin\zeta\, f^x_{obs}, \nonumber \\[2mm]
E^y_{obs} &=& \delta^2\, l_p^{1/2}\, \sin\zeta\, f^y_{obs}, \nonumber  \\[3mm]
|P| &=& (E^x_{obs})^2 + (E^y_{obs})^2,
\end{eqnarray}
where $f^x_{obs}$ and $f^y_{obs}$ are given by eq. ($\ref{obs_polarization}$).

The direction of the polarization vector in the observer's frame is measured by the Electric Vector Position Angle (EVPA) defined as

\begin{eqnarray}
EVPA = \arctan\left({-\frac{f^x_{obs}}{f^y_{obs}}}\right).
\end{eqnarray}
In our simulations we estimate the EVPA counter-clockwise with respect to the positive semi-axis $x>0$ on the observer's sky.


\section{Linear polarization in static wormhole spacetime}

In this section we apply the model of a magnetized fluid ring emitting synchrotron radiation to obtain the observable polarization in static wormhole spacetime. Using a ray-tracing procedure we integrate numerically the null geodesic equations for trajectories originating from a particular equatorial circular orbit with radius $r_s$. As a result, each trajectory projects on the observer's sky with celestial coordinates $x$ and $y$. The emission model in the fluid rest frame determines the constants of motion $\kappa_1$ and $\kappa_2$ on the photon trajectory in terms of the components of the polarization vector and the 4-velocity in the local frame as expressed by eq. ($\ref{kappa_rfr}$). Then, the apparent polarization at each point of the observer's sky with coordinates $(x,y)$ is completely defined by the values of integrals of motion $\kappa_1$ and $\kappa_2$ according to eq. ($\ref{obs_polarization}$). We further calculate the polarization intensity and direction EVPA, which represent the experimentally  observable quantities. In our images the polarization intensity is proportional to the length of the polarization vector while EVPA is determined by its tilt with respect to the positive direction of the $x$-axis.

The images of the circular orbits can be classified into direct and indirect ones according to the type of the photon trajectories which lead to their formation. Direct images arise from photon trajectories which are characterized by the variation of the azimuthal angle in the range  $\phi\in[0,\pi)$. On the other hand, for the indirect images the azimuthal angle varies in the interval $\phi\in[0, (k+1)\pi]$, where the positive integer $k$ determines the order of the image. This implies that the photon trajectories which form indirect images of order $k$ perform $k$ half-loops around the compact object before reaching the observer. In our study we will consider both direct images and indirect images of order $k=1$.

In our simplified model of a magnetized fluid ring orbiting around the central compact object the observable polarization depends on a range of physical parameters. On the one hand we have the characteristics of the accretion flux, which include the direction of the magnetic field and the fluid velocity. On the other hand the spacetime geometry affects the images by means of the parameters of the metric, which describes the compact object, the inclination angle of the observer, and the emission radius. Since we would like to probe the hypothesis that an exotic compact object may lead to a similar polarization distribution as observed for M87*, we give preference to physical parameters which reproduce qualitatively the EHT polarimetric observations. For this reason we consider small inclination angles close to the estimated inclination angle $\theta=17^\circ$ for the galaxy M87*. The fluid is assumed to move inwards which is the natural direction of the accretion flow, and it rotates in a clockwise direction as observed for M87*.  We further pick emission orbits located in the region of strong gravitational field in the vicinity of the ISCO. In particular, we consider the ISCO for the Schwarzschild solution and the orbit with radius $r=4.5M$ which corresponds to the apparent size of the emission ring in the EHT observations of M87*.


We further put constraints on the local magnetic field by taking advantage of the insights provided by previous studies \cite{Narayan:2021}.  Modelling  the polarization of M87* by means of an equatorial synchrotron emission in the Schwarzschild background, it was demonstrated  that vertical magnetic field leads to a very different polarization pattern than the experimentally detected. The asymmetry of the observed polarization intensity along the ring is controlled by the physical effects of Doppler beaming and abberation. For vertical magnetic field and in-falling fluid, which rotates in the clockwise direction the two effects compete and favor the increasing of the polarization intensity at the opposite sides of the image. The effect of the abberation dominates and causes that the largest intensity is distributed in the left or down part of the images depending on the direction of the fluid velocity. Similar pattern is observed if we consider only the influence of the strong gravitational field on the vertically magnetized fluid assuming that the fluid is at rest. The gravitational lensing leads to maximum observable intensity of the polarized flux concentrated at the bottom of the ring.

This pattern contradicts the observations by the EHT since the image of M87* is characterized by a strong linearly polarized flux on its right-hand side which gradually shades down towards the north and south poles. Such a distribution of the polarization intensity can be achieved by restricting the magnetic field to the equatorial plane. If the magnetic field contains only equatorial components, the Doppler beaming and the abberation lead to the same type of flux asymmmetry and enhance each other. As a result,  the polarization intensity concentrates on the right-hand side of the image resembling the observed pattern.

Purely radial or azimuthal magnetic fields are not suitable for modeling M87* since they don't reproduce the twist of the polarization vector. They lead to polarization in purely tangential or radial direction, respectively, while the observed image is characterized by a twisting polarization pattern where the polarization vector rotates up to $70^\circ$  around the ring, measured with respect to the radial direction. Thus, it was estimated in \cite{Narayan:2021} that assuming that the spacetime geometry  is described by the Schwarzschild solution the emission from a magnetized fluid ring reproduces the observed polarization pattern most accurately if the magnetic field is purely equatorial with radial component dominating over the azimuthal one. Probing several directions of the magnetic field it was suggested that a magnetic field aligned as $B= [B_r, B_\phi, B_\theta] = [0.87, 0.5, 0]$  describes best the observed polarization of M87*.

We can consider that the assumption of a purely equatorial magnetic field already determines the direction of the fluid velocity. In the inner region of the disk it is most probable that the magnetic field follows the flow of the fluid velocity, i.e. the angles $\eta$ and $\xi$  are related as  $\eta =\xi$ or $\eta = \pi + \xi$. The two choices correspond to parallel magnetic and velocity fields with coinciding or opposite directions as the ambiguity does not influence the polarization pattern.

In the following we investigate how the polarization properties of the fluid ring model are influenced by the spacetime geometry by performing a range of simulations in static wormhole spacetime for various inclination angles and magnetic field directions. We consider  directly lensed images as well as indirect images of order $k=1$, which form by photon trajectories making one half-turn around the compact object before reaching the observer.

 \subsection {Direct images}

As a first step we study the polarization in vertical magnetic field. In particular, we are interested in examining whether vertical magnetic field in wormhole spacetime leads to a polarization picture which is inconsistent with the observations of M87* similar to the conclusion for the Schwarzschild black hole. The effect of abberation which  causes increased polarization intensity in the left-hand side of the image is  kinematic one and it will appear in any static spacetime. We should investigate however the influence of the gravitational lensing in the region of strong gravitational field and ensure that it does not have the opposite impact, so that it  compensates the abberation and reverses the asymmetry in the observable polarization intensity.

For the purpose we simulate the observable polarization produced by a fluid ring at rest with purely vertical magnetic field. In order to probe the strong gravitational interaction we choose two values of the ring radius given by $r=6M$ and $r=4.5M$. We vary the redshift function of the wormhole geometry in the limits $\alpha\in[0,3]$ and compare the apparent linear polarization with the emission in the spacetime of the Schwarzschild black hole under the same conditions.

\begin{figure}[t]
\centering
   \includegraphics[width=0.58\textwidth]{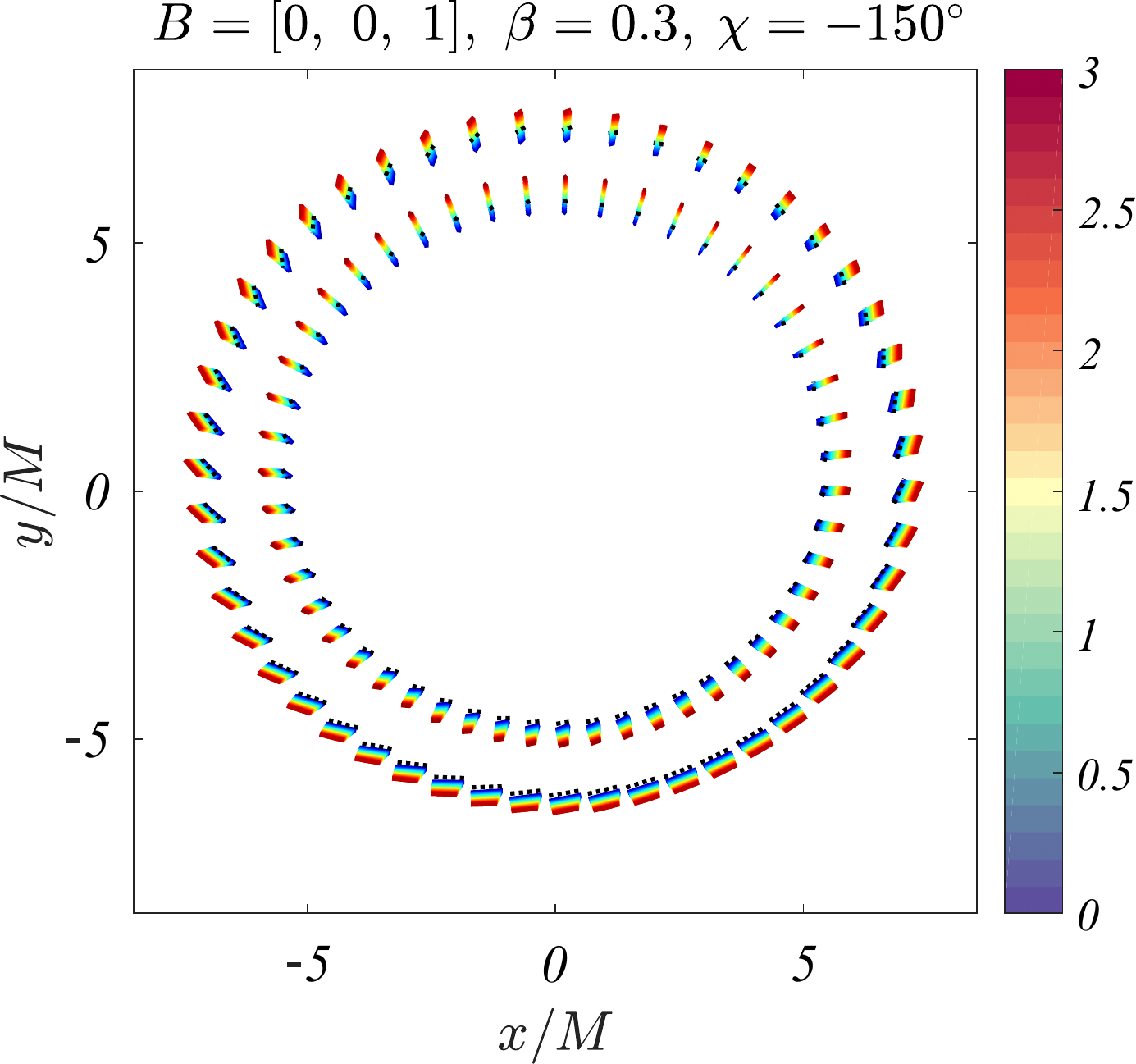}
 \caption{Polarization in vertical magnetic field for wormholes with different redshift parameter $\alpha$. Each color represents the observable polarization of the orbits located at $r=6M$ (outer ring) and $r=4.5M$ (inner ring) for a particular wormhole solution with  $\alpha\in[0,3]$. The polarization for the Schwarzschild black hole is given by a black dotted line as a reference. The inclination angle is $\theta = 20^\circ$.}
\label{fig:pol_20v}
\end{figure}

In Fig. $\ref{fig:pol_20v}$ we see that the polarization pattern for the wormhole spacetime resembles closely the distribution in the Schwarzschild case. The polarization intensity is smaller than for the Schwarzschild black hole and further decreases when we increase the redshift parameter $\alpha$. However, it varies in the same manner along the ring reaching its maximum values at the bottom part of the image and declining towards the top. Thus, we can conclude that if we describe the spacetime by a  wormhole geometry the emission of a fluid ring in a vertical magnetic field would still leads to contradiction with the observed  polarization of M87* similar to the Schwarzschild black hole. For this reason we focus in our analysis on purely equatorial magnetic fields.

In order to get intuition about the variation of the observable polarization due to the different spacetime geometry we plot the polarized image of the rings located at $r=6M$ and $r=4.5M$ for a continuous distribution of the redshift parameter $\alpha$ and several representative configurations of equatorial magnetic field.



\begin{figure}[]
\centering
   \includegraphics[width=\textwidth]{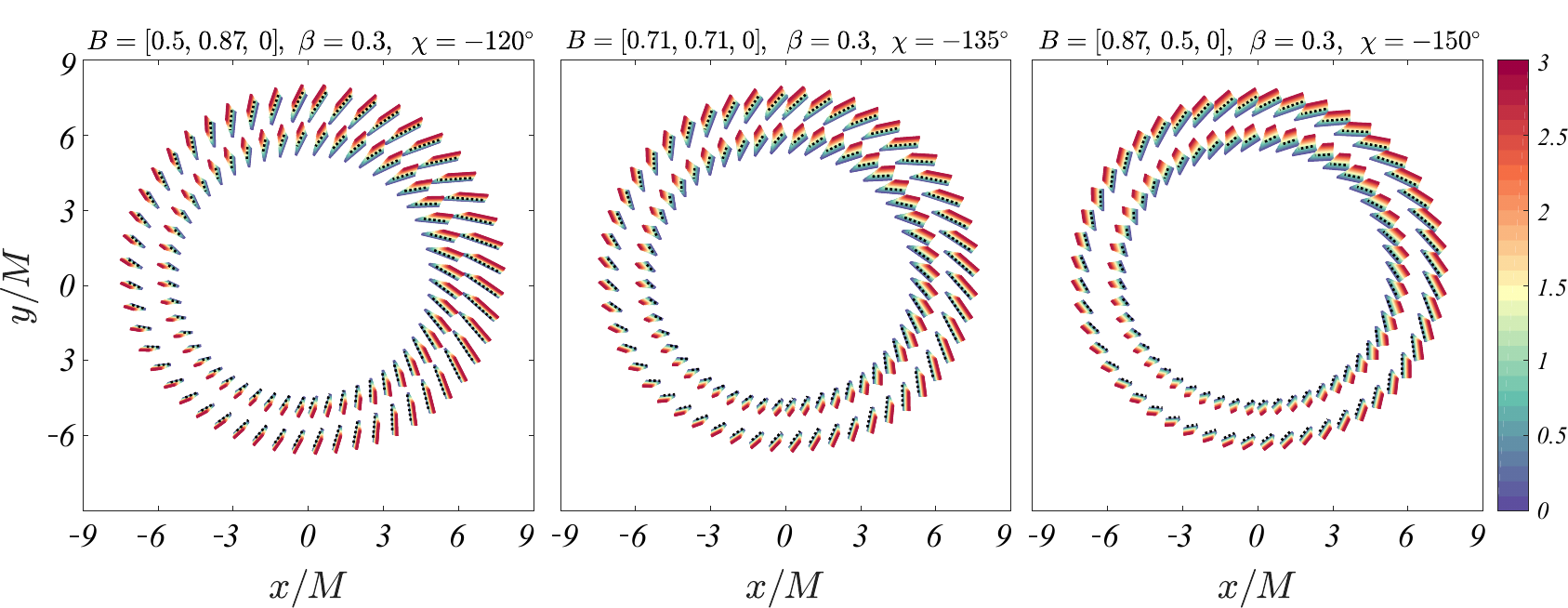}
 \caption{Polarization in equatorial magnetic field for wormholes with different redshift parameter $\alpha$. Each color represents the observable polarization of the orbits located at $r=6M$ (outer ring) and $r=4.5M$ (inner ring) for a particular wormhole solution with  $\alpha\in[0,3]$. The polarization for the Schwarzschild black hole is given by a black dotted line as a reference. The inclination angle is $\theta = 20^\circ$.}
\label{fig:pol_20}
\end{figure}

We observe that the polarization in the wormhole spacetime obeys a qualitatively similar distribution as for the Schwarzschild black hole. The polarization intensity increases with respect to the Schwarzschild case for small $\alpha$ and decreases when the redschift parameter grows. However, its relative distribution around the ring follows the same pattern as for the Schwarzschild black hole. In Fig. $\ref{fig:pol_20}$ we see that the maximum intensity of the polarized flux is again concentrated at the right-hand side of the image and decays in a similar fashion towards the north and south poles. The direction of the polarization vector deviates very slightly from the Schwarzschild solution for all the values of the azimuthal coordinate and the redshift parameter. As a result it reproduces a similar twist around the ring.

From these arguments we can conclude that the direct gravitational lensing around wormholes can lead to a similar polarization picture as for black holes if we consider  small inclination angles. Thus, the direct equatorial emission in wormhole spacetimes can reproduce the polarization data for M87* in a comparable way as the corresponding model in the Schwarzschild spacetime.

In order to assess quantitatively the variation in the polarization picture for the two types of compact objects we  analyse the deviation of the polarization intensity and its direction. Since we would like to investigate whether we can distinguish wormholes and black holes observationally by means of polarization measurements, we compare the features of the linear polarization for the same apparent radius rather than for the same emitting one. Due to the different degree of focusing resulting from the gravitational lensing, the polarized flux which we consider originates from different emission radii in the two spacetimes. However, if we restrict ourselves to direct images, the deviation in the emitting radius is not substantial (see Appendix A).

In Fig. $\ref{fig:polarization_20}$ we present the analysis of the relative polarization properties of the Schwarzschild black hole and the class of static traversable wormholes with different redshift parameter $\alpha$. For each $\alpha$ we consider the polarization at the observable location of the ISCO for the Schwarzschild black hole. In this way we can explore how the polarization properties at a given point of the celestial sphere will be modified if we vary the spacetime geometry. The inclination angle is $\theta = 20^\circ$ and the magnetic field is purely equatorial.

\begin{figure}[t]
\centering
  \includegraphics[width=0.99\textwidth]{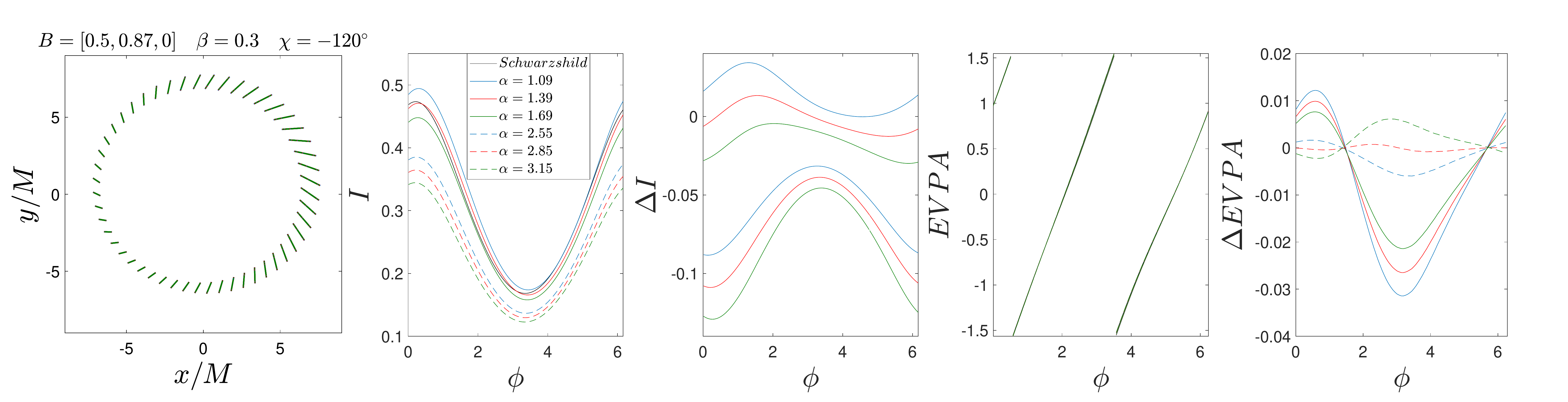}\\[1mm]
  \includegraphics[width=0.99\textwidth]{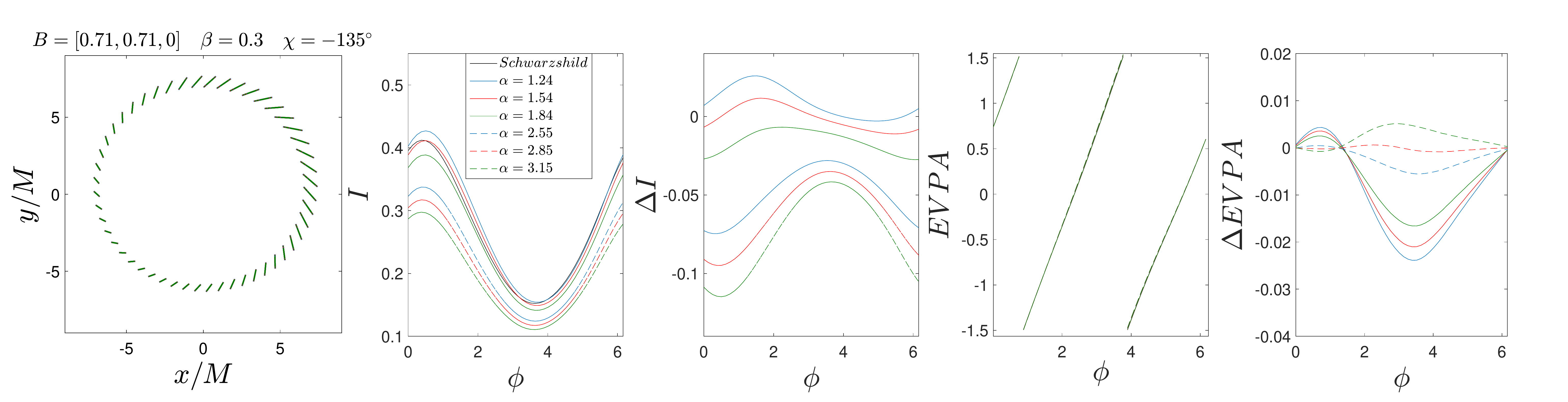}\\[1mm]
  \includegraphics[width=0.99\textwidth]{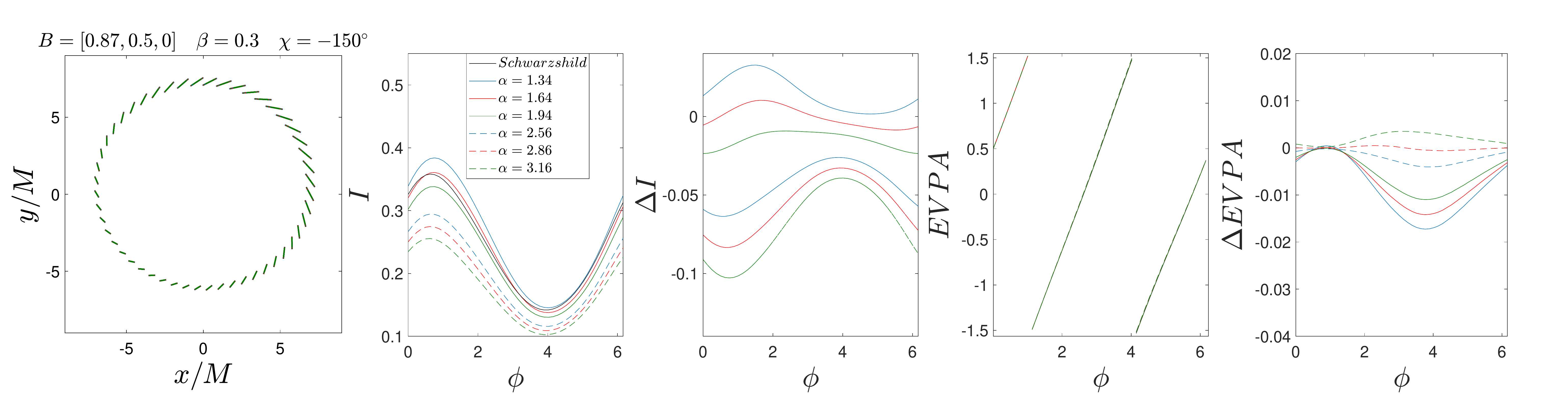}
 \caption{Linear polarization for static wormholes at the inclination angle $\theta =20^\circ$. We analyze the polarization intensity I and direction EVPA as a function of the redshift parameter $\alpha$, as well as their deviation from the Schwarzschild black hole $\Delta$I and  $\Delta$EVPA (see main text). The two critical values of $\alpha$, which lead to a minimal deviation $\Delta$I and  $\Delta$EVPA  are presented in red solid and red dashed lines, respectively. }
\label{fig:polarization_20}
\end{figure}

For each image we plot the variation of the polarization intensity and its direction around the ring as a function of the azimuthal coordinate. We further present their deviation from the Schwarzschild solution defining the relative intensity $\Delta \text{I} = \text{I}_\text{WH} - \text{I}_\text{Sch}$ and the relative direction angle $\Delta \text{EVPA} =\text{ EVPA}_\text{WH} - \text{EVPA}_\text{Sch}$ at each point of the image. Examining these quantities we can reach the following conclusions. For each $\alpha$ the variation of the  intensity and the direction angle as a function of the azimuthal angle possesses a similar profile as for the Schwarzschild solution. The deviation between the wormhole and the black hole polarization characteristics is not substantial for all the values of $\alpha\in[0,3]$. Furthermore, we notice that for small values of the redshift parameter the polarization intensity in the wormhole spacetime is larger than that for the Schwarzschild black hole at each point of the image while for larger $\alpha$ the opposite inequality applies. This implies that there exists some critical  value of the redshift parameter $\alpha^{(1)}_{crit}$, for which the deviation in the polarization intensity for the two types of geometries is minimal.

The behavior of the polarization direction angle possesses a similar feature. For small values of  $\alpha$ the deviation $\Delta$ EVPA is positive in some region at the right-hand side of the image and negative for the other azimuthal angles while for large redshift parameter we observe the opposite trend. Thus, we can conclude that there exists a second critical value $\alpha^{(2)}_{crit}$, for which the direction of the polarization vector for the wormhole solution deviates minimally from the Schwarzschild black hole.

\begin{figure}[]
\centering
 \subfloat[][]{\includegraphics[width=0.75\textwidth]{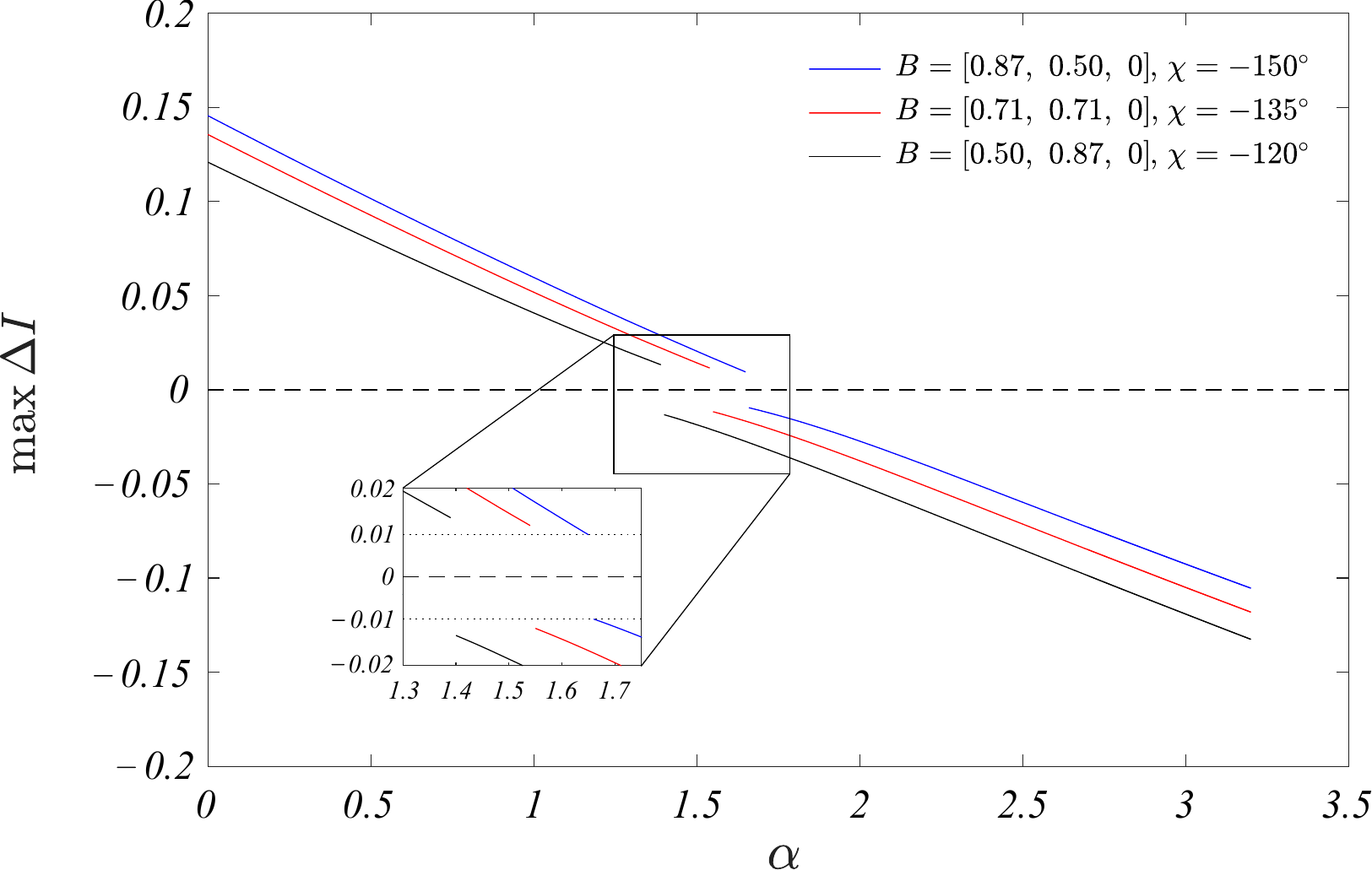}} \\[2mm]
 \subfloat[][]{\includegraphics[width=0.75\textwidth]{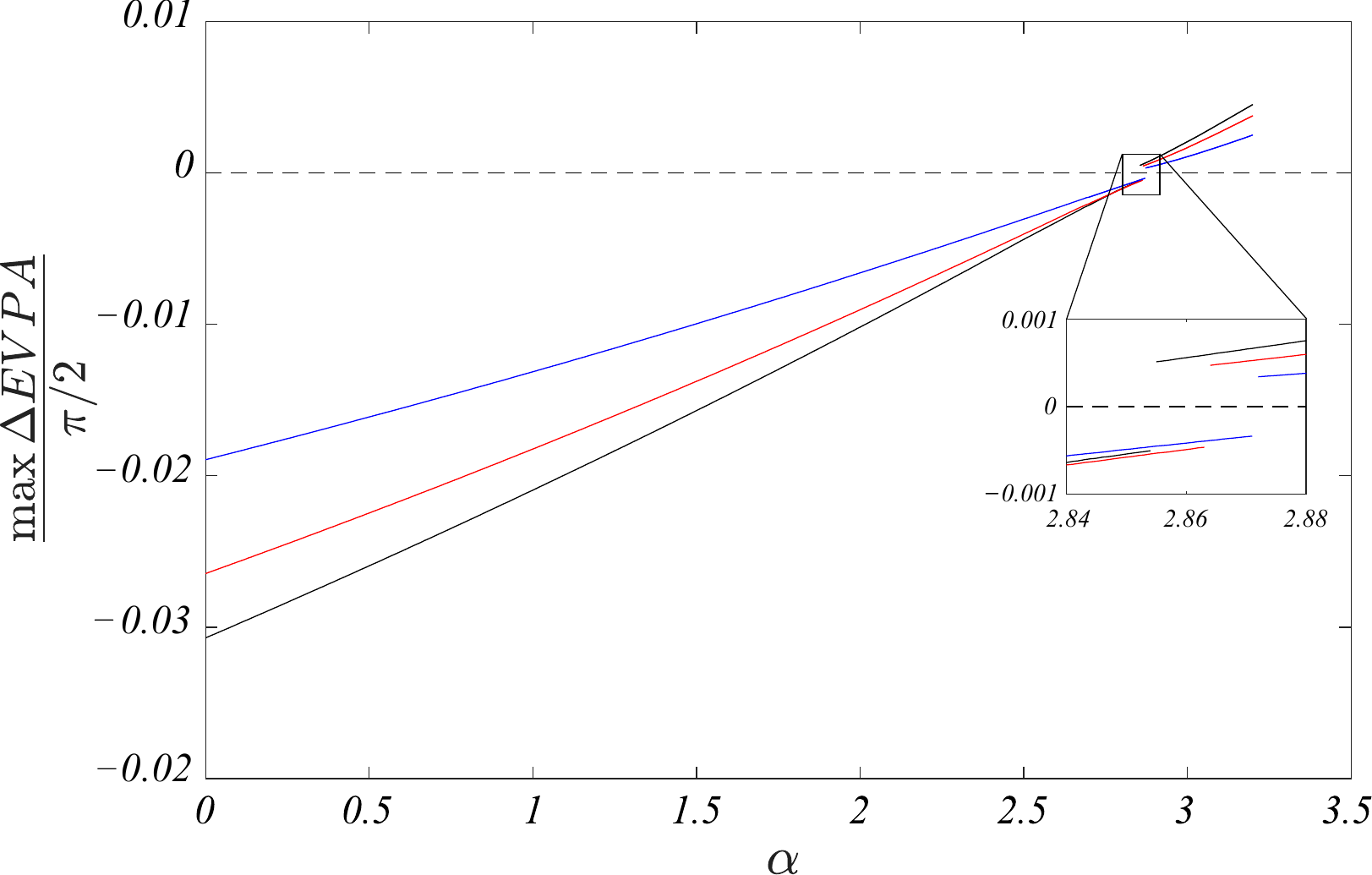}}
\caption{Maximum deviation of the polarization properties of the wormhole from the Schwarzschild black hole as a function of the redshift parameter $\alpha$. We plot the maximum deviation of the intensity max $\Delta$I (a), and the polarization direction max $\Delta$EVPA (b) which is reached in the images presented in Fig. $\ref{fig:polarization_20}$  for each value of $\alpha$. Negative values imply that the corresponding quantity is larger for the Schwarzschild solution than for the wormhole.}
\label{fig:maxI_20}
\end{figure}

In Fig. $\ref{fig:maxI_20}$ we examine in more details the deviation of the intensity $\Delta$I and the polarization angle $\Delta$EVPA in the wormhole geometry with respect to the Schwarzschild solution. Since these quantities vary depending on the azimuthal angle around the ring we estimate the maximum deviation of each of them, which can be reached in the image for a fixed value of the redshift parameter and plot it as a function of $\alpha$. In this way we  can measure the upper limit of the possible discrepancy in the polarization properties of the two geometries for any redshift parameter. From this analysis we can deduce the two critical values of $\alpha$, which lead to minimal deviation in the polarization intensity and direction, respectively, and estimate its amount.

The results are presented in Table $\ref{table:theta20}$ for the three configurations of equatorial magnetic field, which we consider. Thus, for $B = [0.87, 0.5, 0]$ we obtain that the wormhole solution with $\alpha^{(1)}_{crit}= 1.64$ possesses polarization intensity whose distribution resembles most closely the Schwarzschild black hole deviating less than $3.3\, \%$ at any point of the image. For the same value of the redshift parameter the polarization direction deviates less than $1.1\,\%$. On the other hand, the wormhole solution for $\alpha^{(2)}_{crit}= 2.86$ resembles most closely the Schwarzschild black hole in the polarization twist around the ring with deviation of the direction vector less than $0.04\,\%$. For the same value of $\alpha$ the polarization direction deviates less than $23.4\,\%$ at any point of the image. The values for the other magnetic field configurations are interpreted in a similar way.

\begin{table}[t!]
    \centering
      \begin{tabular}{||c|c|c|c|c|c||}
       \hline
         \thead{ Magnetic field }   & \thead {Minimization \\ criteria} & \thead{$\frac{\text{max}\,\Delta \text{I}}{\text{I}_\text{Sch}} \,\, [\%]$}
          &\thead{ $\frac{\text{max}\,\Delta \text{EVPA}}{\text{EVPA}_\text{Sch}}\,\, [\%]$}&  $\phi \, [rad]$ & $\alpha_{crit}$   \\    \hline
           \thead{$\text{B = [0.5, 0.87, 0]}$}  & \thead{$\Delta$I \\[2mm] $\Delta$EVPA} & \thead{3.8 \\[2mm] 23.0} & \thead{2.2 \\[2mm] 0.3} & \thead{$0.48\pi$ \\[2mm] $0.73\pi$ } &\thead{1.39\\[2mm]2.85}
          \\  \hline
          \thead{$\text{B= [0.71, 0.71, 0]}$ } & \thead{$\Delta$I \\[2mm] $\Delta$EVPA} & \thead{3.6 \\[2mm] 23.1} & \thead{1.8 \\[2mm] 0.07} &  \thead{$0.53\pi$ \\[2mm] $1.32\pi$ } &\thead{1.54\\[2mm]2.85}
          \\  \hline
           \thead{$\text{B= [0.87, 0.5, 0]}$}  & \thead{$\Delta$I \\[2mm] $\Delta$EVPA} & \thead{3.3 \\[2mm] 23.4} & \thead{1.1 \\[2mm] 0.04} &  \thead{$0.53\pi$ \\[2mm] $0.32\pi$ } &\thead{1.64\\[2mm]2.86}
          \\  \hline
       \end{tabular}
      \caption{Deviation of the wormhole polarization from the Schwarzschild black hole for the critical values of the redshift parameter $\alpha$, for which $\Delta$I or $\Delta$EVPA are minimal. In each case we give  the maximum relative deviations $\text{max}\,\Delta\text{I}/\text{I}_\text{Sch}$ and $\text{max}\,\Delta\text{EVPA}/\text{EVPA}_\text{Sch}$   with respect to the  Schwarzschild solution, which are reached in the polarized images and the corresponding azimuthal angle. }
    \label{table:theta20}
\end{table}


Examining the particular form of the functions max $\Delta$I($\alpha$) and max $\Delta$EVPA($\alpha$) we see that the wormhole solution with the minimal deviation $\Delta I$ can be considered with a good precision to  possesses the most similar polarization structure to the Schwarzschild black hole. Thus, we can assume that for $\alpha=\alpha^{(1)}_{crit}$ the polarization of the wormhole resembles most closely the Schwarzschild solution taking into account both its intensity and twist.

In Fig. $\ref{fig:maxI_20}$ we obtain also  an estimate for the maximal deviation from the polarization properties of the Schwarzschild black hole which is possible for our class of wormholes. We see that the wormhole solutions are most clearly distinguished  for small values of $\alpha$ as the discrepancy from the Schwarzschild solution depends on the magnetic field. For example, in the magnetic field configuration $B = [0.87, 0.5, 0]$ we get  for $\alpha =0$ that the maximal deviation in the intensity is $\text{max}\,\Delta \text{I}/\text{I}_\text{Sch} = 43\%$  and the maximal deviation in the polarization angle is $\text{max}\,\Delta \text{EVPA}/\text{EVPA}_\text{Sch} = 2.3\%$ , while for  $B = [0.5, 0.87, 0]$ we get the values $\text{max}\,\Delta \text{I}/\text{I}_\text{Sch} = 28\%$ and $\text{max}\,\Delta \text{EVPA}/\text{EVPA}_\text{Sch} = 3.8\%$.

The minimal deviation from the Schwarzschild solution, which is possible for every magnetic field configuration, depends insignificantly in magnitude on the magnetic field direction but it corresponds to a different value of the redshift parameter $\alpha$.

The deviation from the Schwarzschild solution for a fixed value of $\alpha$ depends on the magnetic field configuration. For $\alpha < \alpha^{(1)}_{crit}$  $\Delta I$ increases when the radial component $B_r$ grows, while for $\alpha >\alpha^{(1)}_{crit}$ we observer the opposite correlation. The deviation of the polarization angle increases when $B_r$ declines for all the values of the redshift parameter which we consider.

We reach similar conclusions if we vary the radius of the emitting ring in the region of strong interaction. In Appendix B we study the polarization observed at location of the image of the orbit with radius $r=4.5M$ for the Schwarzschild black hole. The polarization pattern reproduces qualitatively the characteristics which we described in the case of $r=6M$. The minimal deviation from the Schwarzschild solution which can be achieved for this emission radius is slightly larger than in the previous case reaching approximately  $6\%$ for the intensity and $3\%$ for the polarization direction.

\begin{figure}[t]
\centering
\begin{tabular}{cc}
   \includegraphics[width=0.48\textwidth]{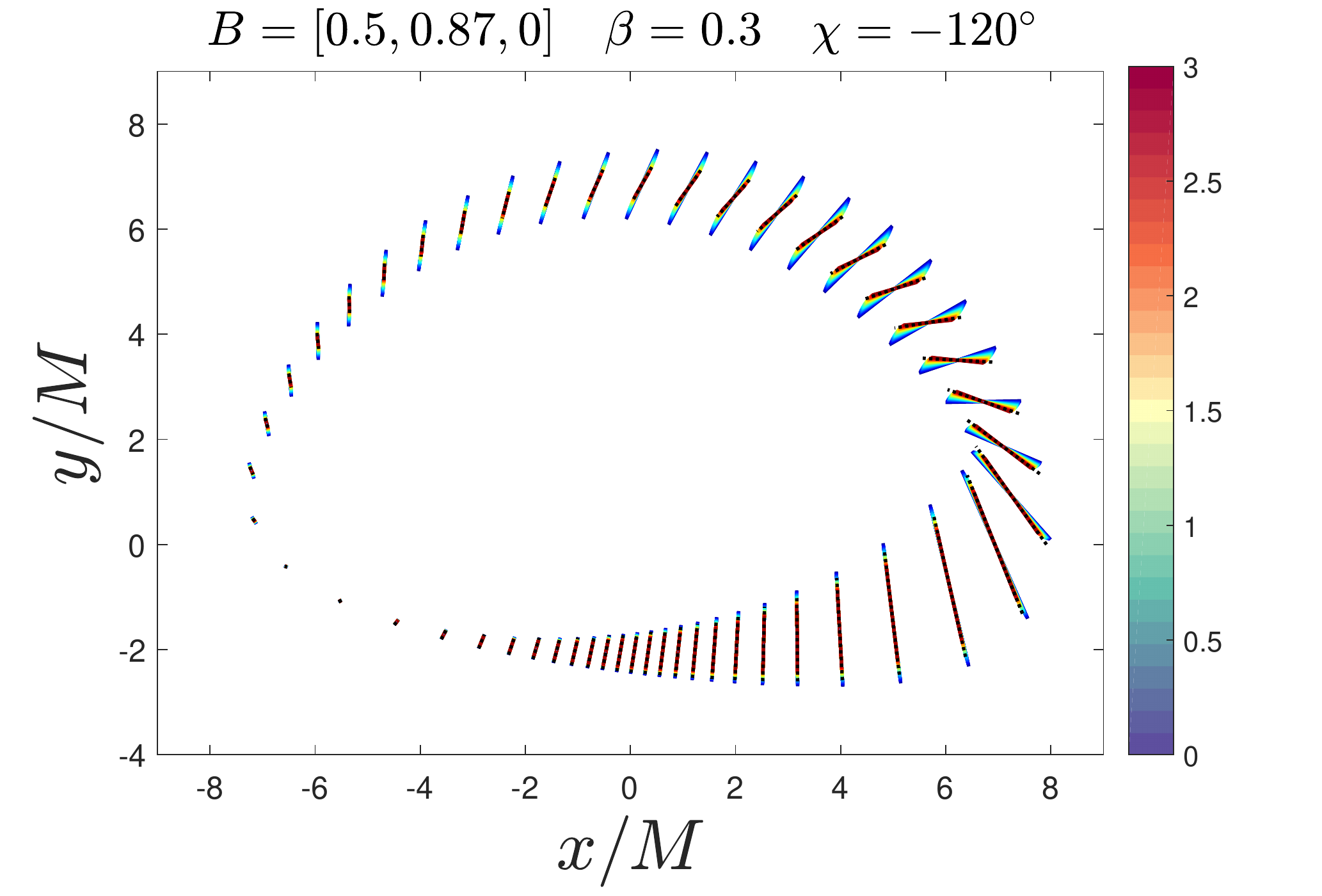}
   \hspace{2mm}
   \includegraphics[width=0.48\textwidth]{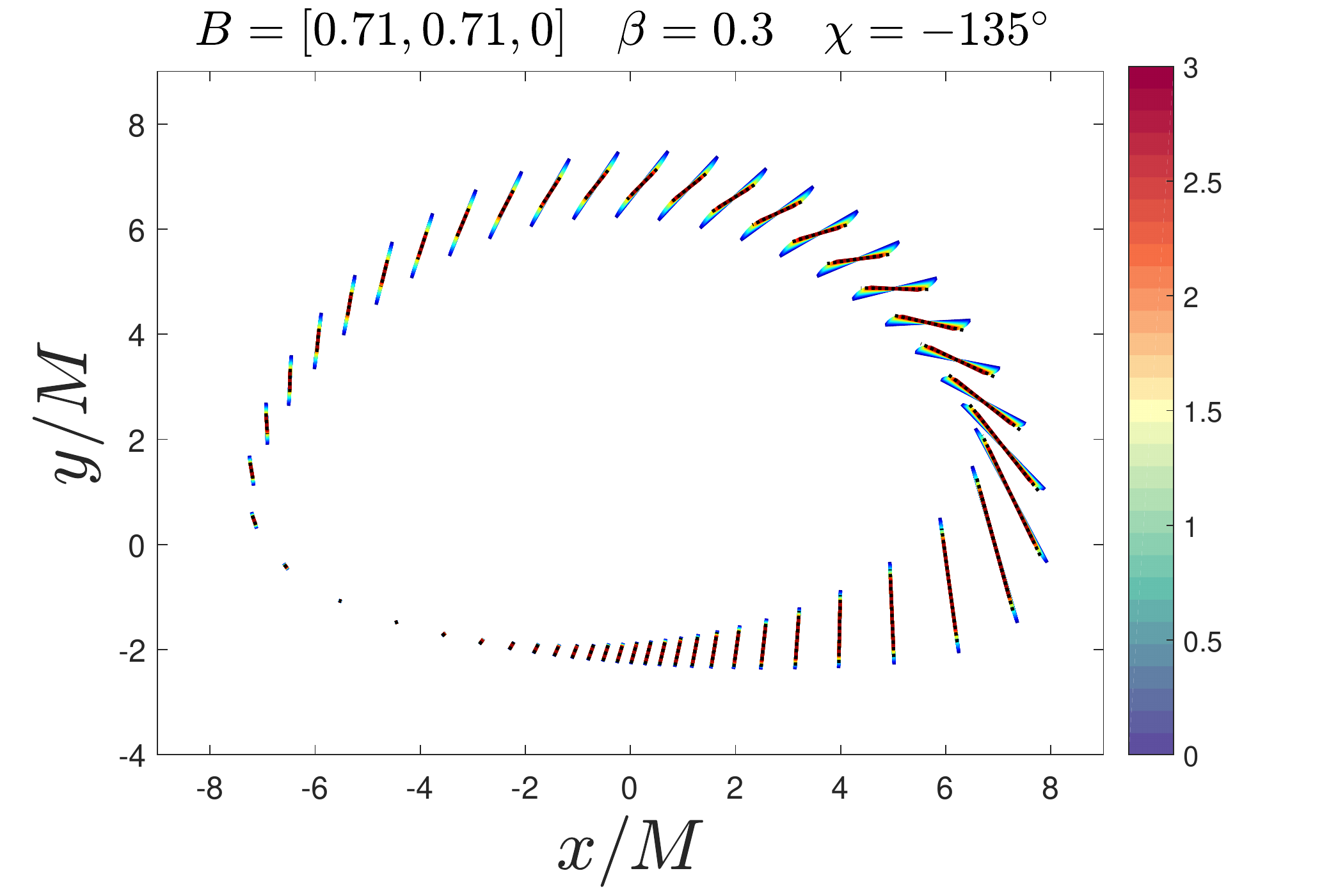}
      \end{tabular}
   \begin{tabular}{c}
   \includegraphics[width=0.48\textwidth]{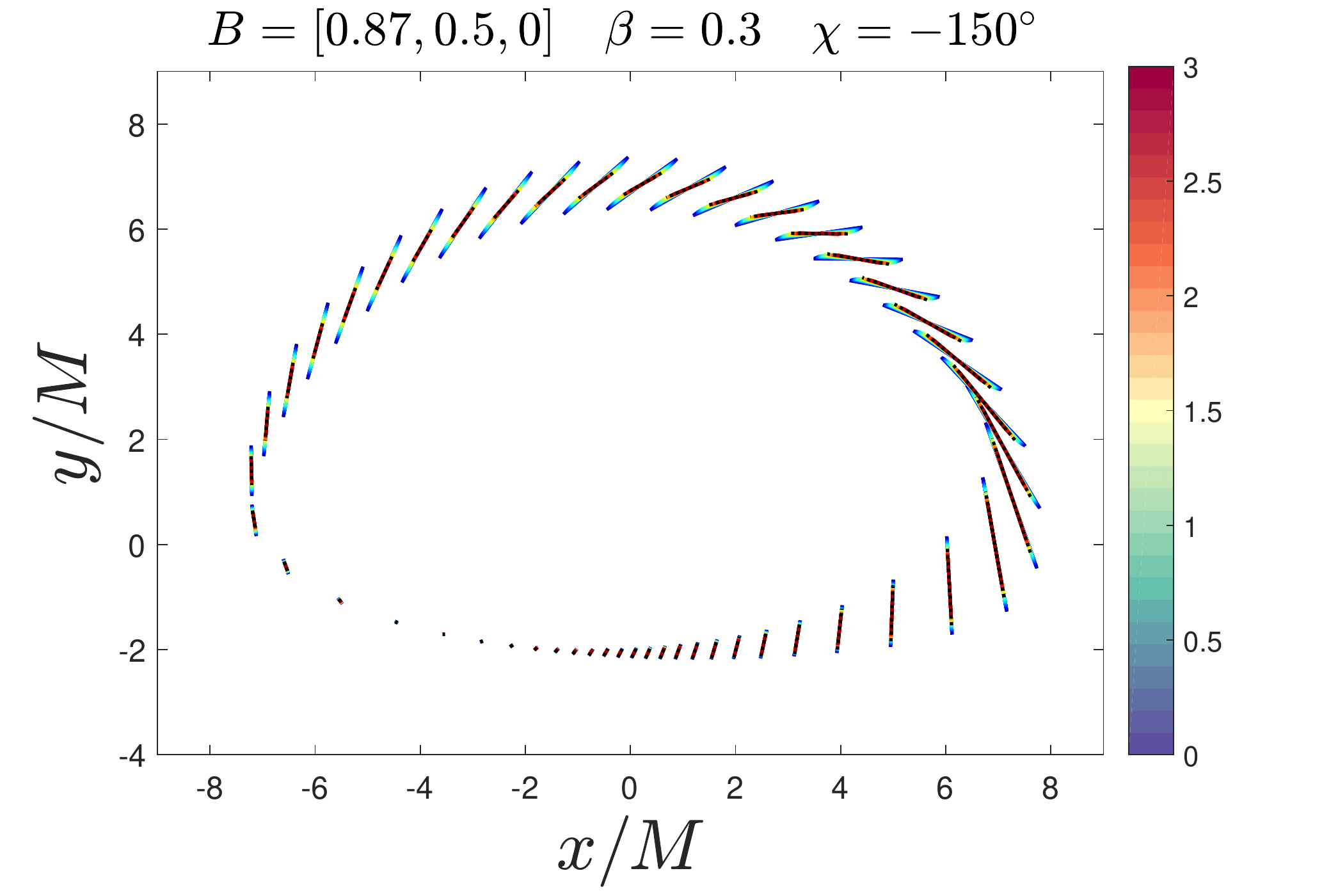}
   \end{tabular}

\caption{Polarization in equatorial magnetic field for wormholes with different redshift parameter $\alpha$. Each color represents the polarization  for a particular wormhole solution with  $\alpha\in[0,3]$ at the observable location of the ISCO for the Schwarzschild solution. The polarization for the Schwarzschild black hole is given by a black dotted line as a reference. The inclination angle is $\theta = 70^\circ$.}
\label{fig:pol_70}
\end{figure}

\begin{figure}[t]
\centering
   \includegraphics[width=0.99\textwidth]{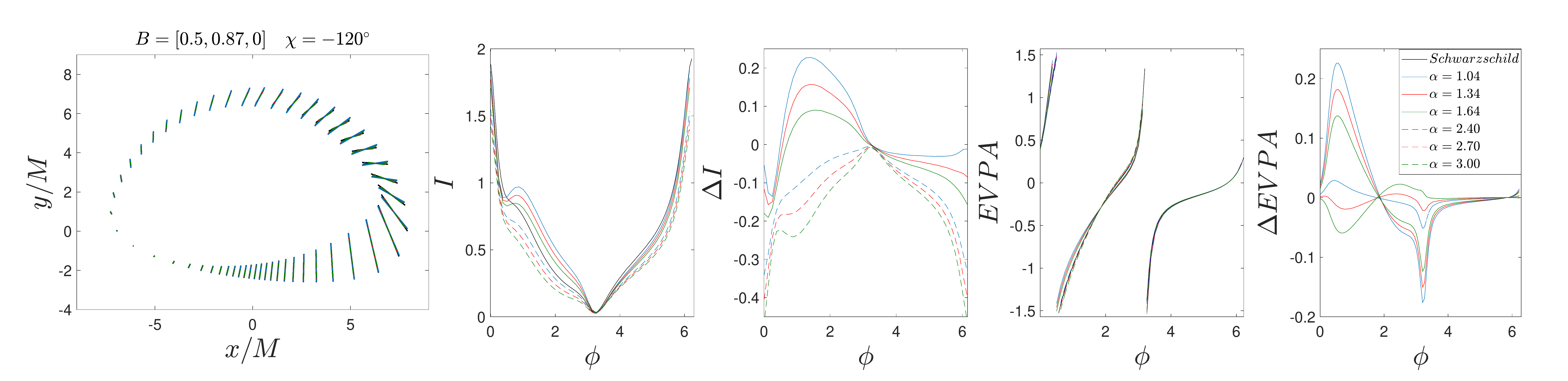}
 \\[2mm]
  \includegraphics[width=0.99\textwidth]{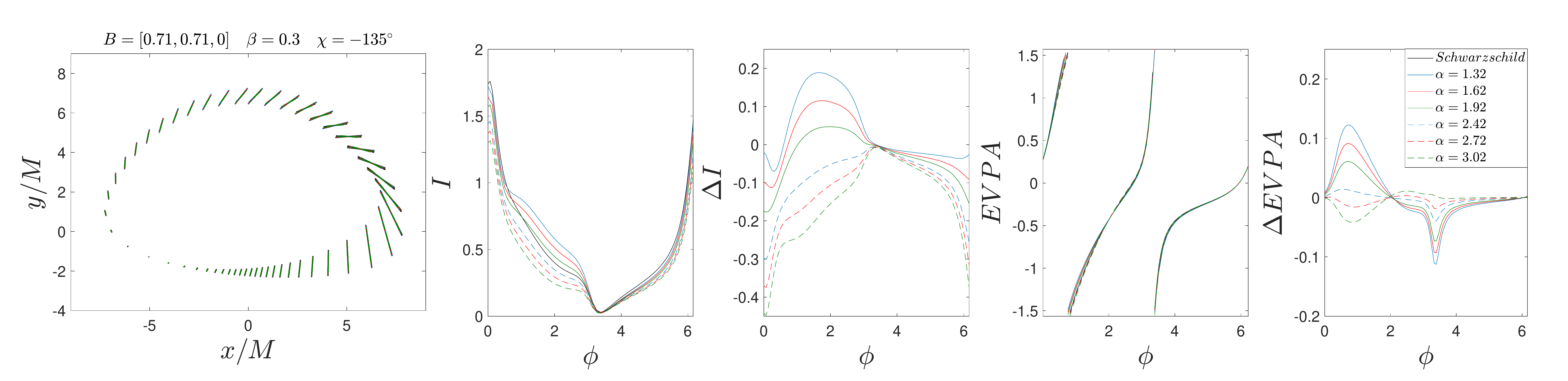} \\[2mm]
  \includegraphics[width=0.99\textwidth]{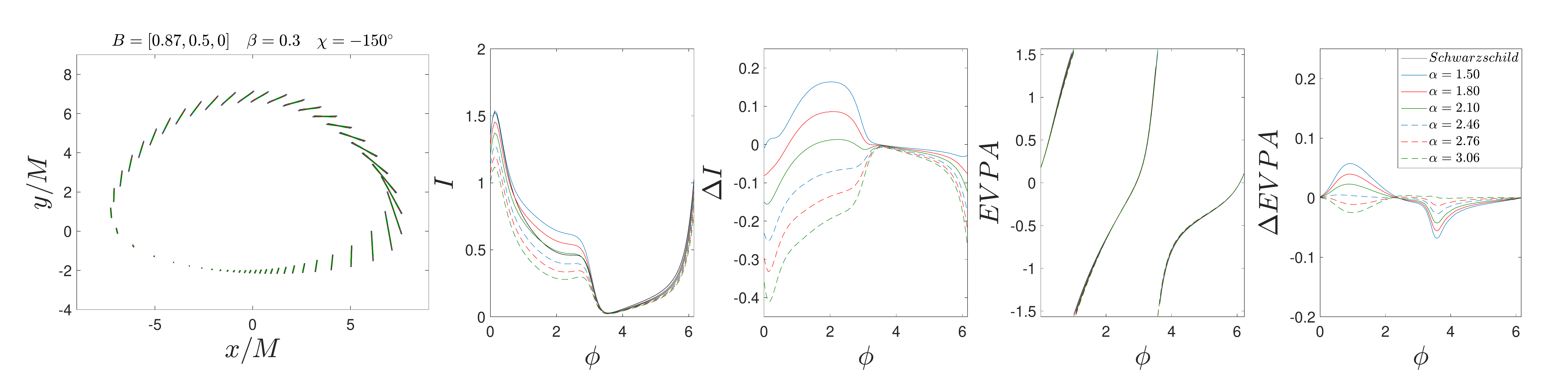}
 \caption{Linear polarization for static wormholes at the inclination angle $\theta =70^\circ$. We analyze the polarization intensity I and direction EVPA as a function of the redshift parameter $\alpha$, as well as their deviation from the Schwarzschild black hole $\Delta$I and  $\Delta$EVPA (see main text). The two critical values of $\alpha$, which lead to a minimal deviation $\Delta$I and  $\Delta$EVPA  are presented in red solid and dashed lines, respectively. }
\label{fig:polarization_70}
\end{figure}

As a next step we consider the influence of the inclination angle on the polarization properties. We obtain the polarized images in the same physical set-up as in the previous examples, however we assume that the disk is tilted at $\theta = 70^\circ$ with respect to the observer. In Fig. $\ref{fig:pol_70}$ we plot the polarization at the observable location of the ISCO for the Schwarzschild solution for a continuous distribution of the redshift parameter $\alpha$. In this way we explore the qualitative features of the polarization pattern for wormhole solutions in comparison to the Schwarzschild black hole. In  Fig. $\ref{fig:polarization_70}$ we further analyse quantitatively the deviation between the polarization properties of the two types of compact objects.

\begin{figure}[]
\centering
 \subfloat[][]{
  \includegraphics[width=0.75\textwidth]{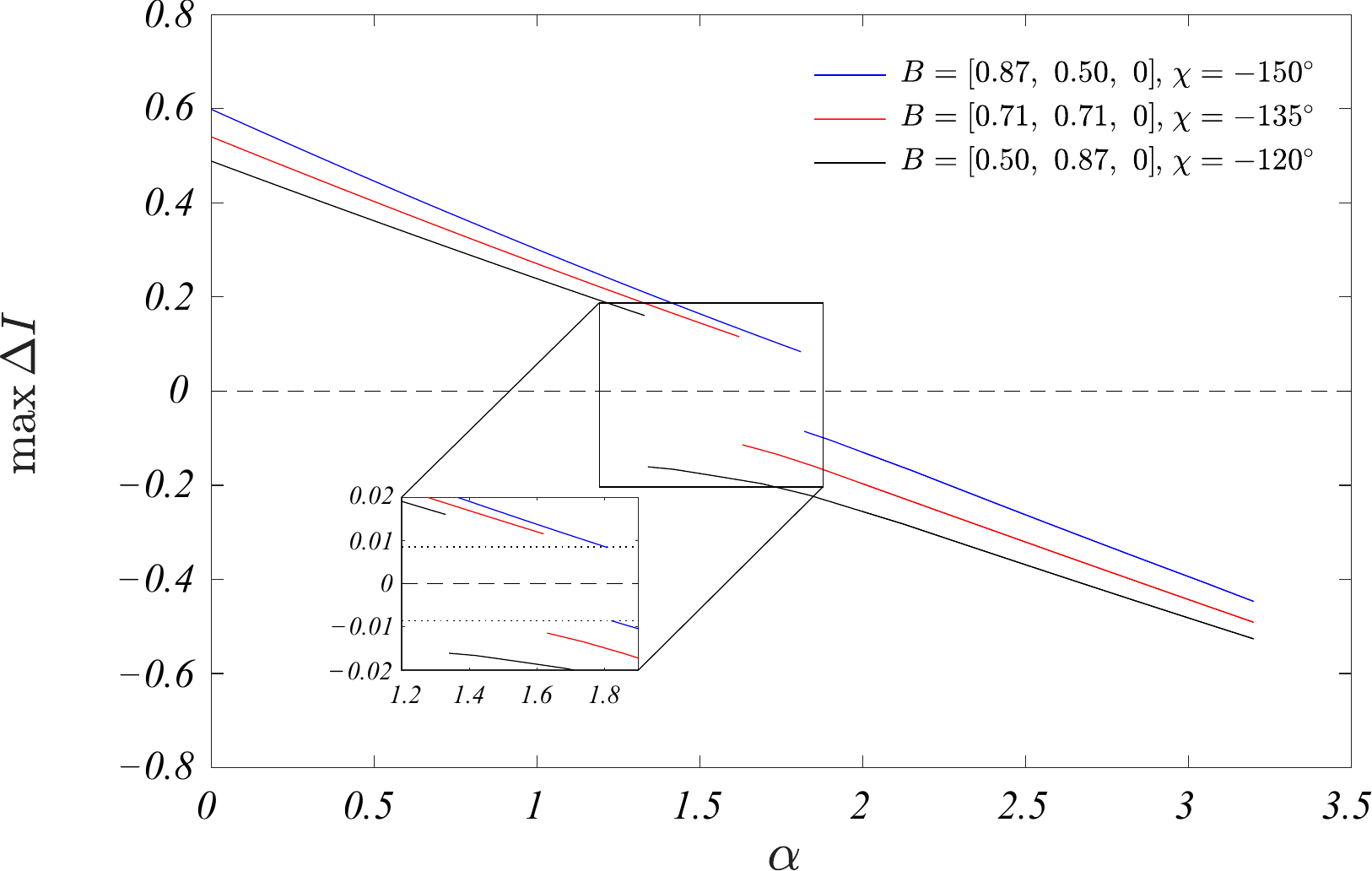} } \\[2mm]
  \subfloat[][]{
  \includegraphics[width=0.75\textwidth]{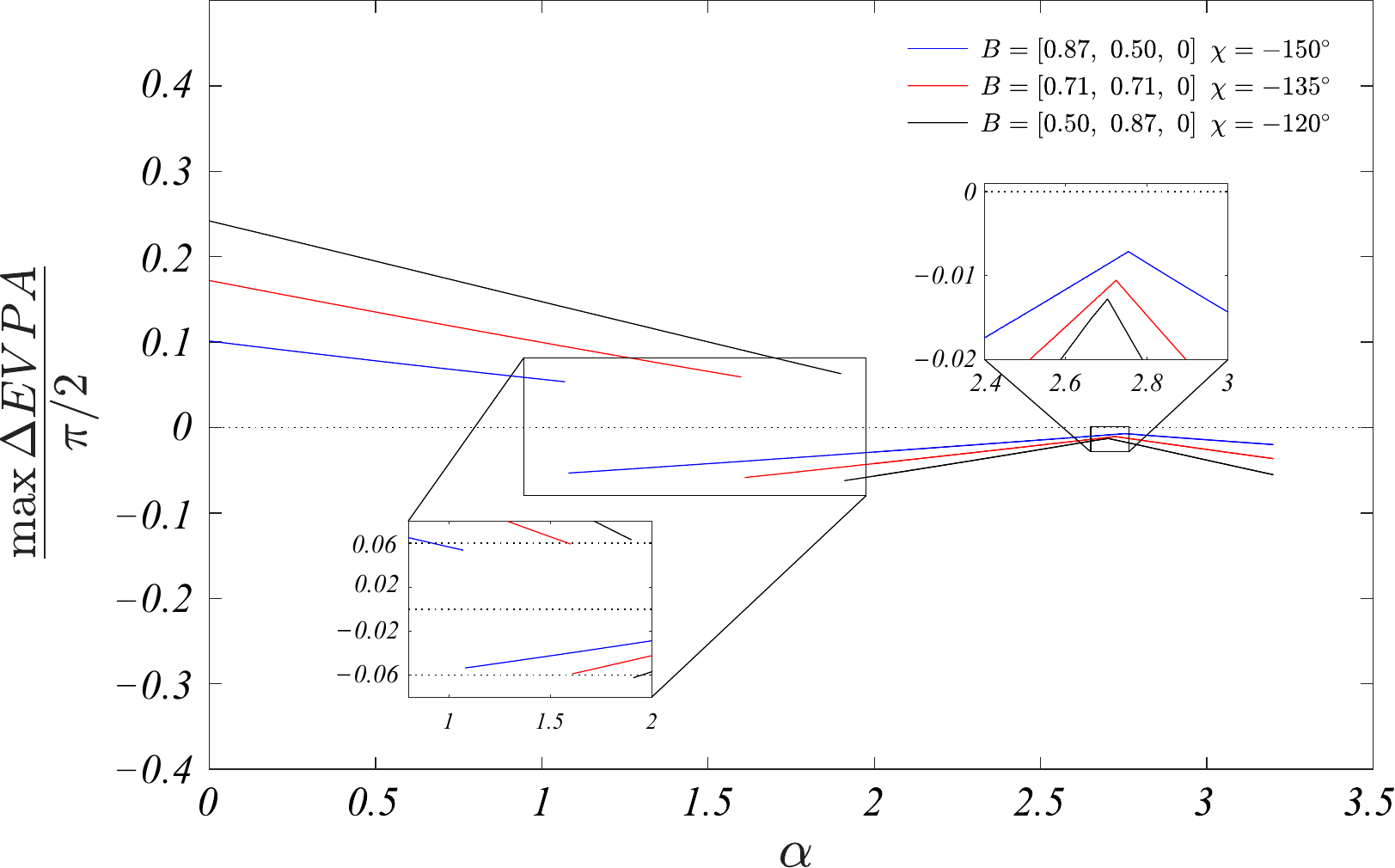}
 }
\caption{Maximum deviation of the polarization properties of the wormhole from the Schwarzschild black hole as a function of the redshift parameter $\alpha$. We plot the maximum deviation of the intensity max $\Delta$I (a), and the polarization direction max $\Delta$EVPA (b) which is reached in the images presented in Fig. $\ref{fig:polarization_70}$  for each value of $\alpha$. Negative values imply that the corresponding quantity is larger for the Schwarzschild solution than for the wormhole.}
\label{fig:maxI_70}
\end{figure}

We see that higher inclination angles lead to more pronounced differences in the polarization properties of the wormhole and black hole geometries. The images possess some common qualitative features as for small inclinations. The polarization pattern is similar to the Schwarzschild black hole with highest intensity at the right-hand side of the image and similar twist around the ring. For small redshift parameter the polarization intensity for the wormhole is higher than  for the Schwarzschild solution, while for larger values of $\alpha$  we observe the opposite relation. The variation of $\Delta$EVPA around the image is more complicated than for small inclination angles. However, we still observe that for smaller $\alpha$ $\Delta$EVPA is positive in some region in the right-hand side of the image and negative for the other azimuthal angles while  for larger $\alpha$  the opposite inequality applies in each of the regions. Thus, there exist a couple of values of the redshift parameter $\alpha^{(1)}_{crit}$ and $\alpha^{(2)}_{crit}$ for  which the polarization intensity or the twist pattern deviate minimally from the Schwarzschild solution, respectively.

In Fig. $\ref{fig:maxI_70}$ we access the deviation from the Schwarzschild black hole for each $\alpha$ by plotting the maximum deviation in the intensity and the EVPA which is reached in the image for the corresponding value of the redshift parameter. Thus, in the case $B = [0.87, 0.5, 0]$ we obtain that for  $\alpha^{(1)}_{crit} = 1.82$ the wormhole polarization intensity resembles most that of the Schwarzschild solution with deviation less than $6.3\%$ , while for  $\alpha^{(2)}_{crit} =2.75$ the polarization twist is most similar with deviation in $\Delta$EVPA less than $0.9\%$. The analysis of the minimal deviation of $\Delta I$ and $\Delta$EVPA for the different equatorial magnetic field configurations is presented in Table $\ref{table:theta70}$.

\begin{table}[t!]
    \centering
      \begin{tabular}{||c|c|c|c|c|c||}
       \hline
         \thead{ Magnetic field }   & \thead {Minimization \\ criteria} & \thead{$\frac{\text{max}\,\Delta \text{I}}{\text{I}_\text{Sch}} \,\, [\%]$}
          &\thead{ $\frac{\text{max}\,\Delta \text{EVPA}}{\text{EVPA}_\text{Sch}}\,\, [\%]$}&  $\phi \, [rad]$ & $\alpha_{crit}$   \\    \hline
           \thead{$\text{B = [0.5, 0.87, 0]}$}  & \thead{$\Delta$I \\[2mm] $\Delta$EVPA} & \thead{12.2 \\[2mm] 21.5} & \thead{11.2 \\[2mm] 1.6} & \thead{$0.06\pi$ \\[2mm] $1.04\pi$ } &\thead{1.33\\[2mm]2.70}
          \\  \hline
          \thead{$\text{B= [0.71, 0.71, 0]}$ } & \thead{$\Delta$I \\[2mm] $\Delta$EVPA} & \thead{7.3 \\[2mm] 21.3} & \thead{6.3 \\[2mm] 1.3} &  \thead{$0.06\pi$ \\[2mm] $1.08\pi$ } &\thead{1.63\\[2mm]2.72}
          \\  \hline
           \thead{$\text{B= [0.87, 0.5, 0]}$}  & \thead{$\Delta$I \\[2mm] $\Delta$EVPA} & \thead{6.3 \\[2mm] 21.5} & \thead{3.6 \\[2mm] 0.9} &  \thead{$1.98\pi$ \\[2mm] $1.14\pi$ } &\thead{1.82\\[2mm]2.75}
          \\  \hline
       \end{tabular}
      \caption{Deviation of the wormhole polarization from the Schwarzschild black hole for the critical values of the redshift parameter $\alpha$, for which $\Delta$I or $\Delta$EVPA are minimal. In each case we give  the maximum relative deviations $\text{max}\,\Delta\text{I}/\text{I}_\text{Sch}$ and $\text{max}\,\Delta\text{EVPA}/\text{EVPA}_\text{Sch}$   with respect to the  Schwarzschild solution, which are reached in the polarized images, and the corresponding azimuthal angle. }
    \label{table:theta70}
\end{table}

\begin{table}[t!]
    \centering
      \begin{tabular}{||c|c|c|c||}
       \hline
         \thead{ Magnetic field }   & \thead {Inclination \\ angle} & \thead{$\frac{\text{max}\,\Delta \text{I}}{\text{I}_\text{Sch}} \,\, $}
          &\thead{ $\frac{\text{max}\,\Delta \text{EVPA}}{\text{EVPA}_\text{Sch}}\,\, $}   \\    \hline
           \thead{$\text{B = [0.5, 0.87, 0]}$}  & \thead{$20^\circ$ \\[2mm] $70^\circ$} & \thead{0.28 \\[2mm] 0.82} & \thead{0.038 \\[2mm] 0.25}
          \\  \hline
           \thead{$\text{B= [0.87, 0.5, 0]}$}  & \thead{$20^\circ$ \\[2mm] $70^\circ$} & \thead{0.43 \\[2mm] 1.28} & \thead{0.023 \\[2mm] 0.11}
          \\  \hline
       \end{tabular}
      \caption{Maximal deviation of the wormhole polarization properties from the Schwarzschild black hole for our class of solutions. The wormhole geometries are most clearly distinguished for the redshift parameter $\alpha=0$.}
    \label{table:maxI}
\end{table}

We see that the minimal discrepancy in the wormhole and black hole polarization increases when the inclination angle grows. The wormholes differ most from the Schwarzschild black hole  for $\alpha =0$ as the deviation depends on the magnetic field. When the radial component $B_r$ increases the difference in the intensity $\Delta I$ grows while $\Delta$EVPA decreases. For  equatorial magnetic fields the maximal discrepancy is significantly larger compared to the results for small inclination angles. For the magnetic field configuration $B = [0.87, 0.5, 0]$ we have approximately 2 times difference in the polarization intensity and  $11\%$ deviation in EVPA. For $B = [0.5, 0.87, 0]$ the maximal deviation for $\alpha=0$ is $\text{max}\,\Delta \text{I}/\text{I}_\text{Sch} = 82\%$ and $\text{max}\,\Delta \text{EVPA}/\text{EVPA}_\text{Sch} = 25\%$. We summarize the data for the maximal discrepancy in the polarization properties with respect to the Schwarzschild solution for small and large inclination angles in Table $\ref{table:maxI}$.

\subsection {Indirect images}

In this section we explore the  polarization resulting from strongly lensed photon trajectories, which perform  $k$ half-loops around the compact object before reaching the observer. In particular, we consider the images of order $k=1$. While the direct photon trajectories exist for a wide range of impact parameters, the values of the impact parameters for the strongly lensed null geodesics are very limited as the interval decreases when the order $k$ grows. As a result, the images of order $k=1$ are restricted to a significantly more compact regions on the observer's sky than the direct images.

\begin{figure}[h!]
\centering
   \includegraphics[width=0.67\textwidth]{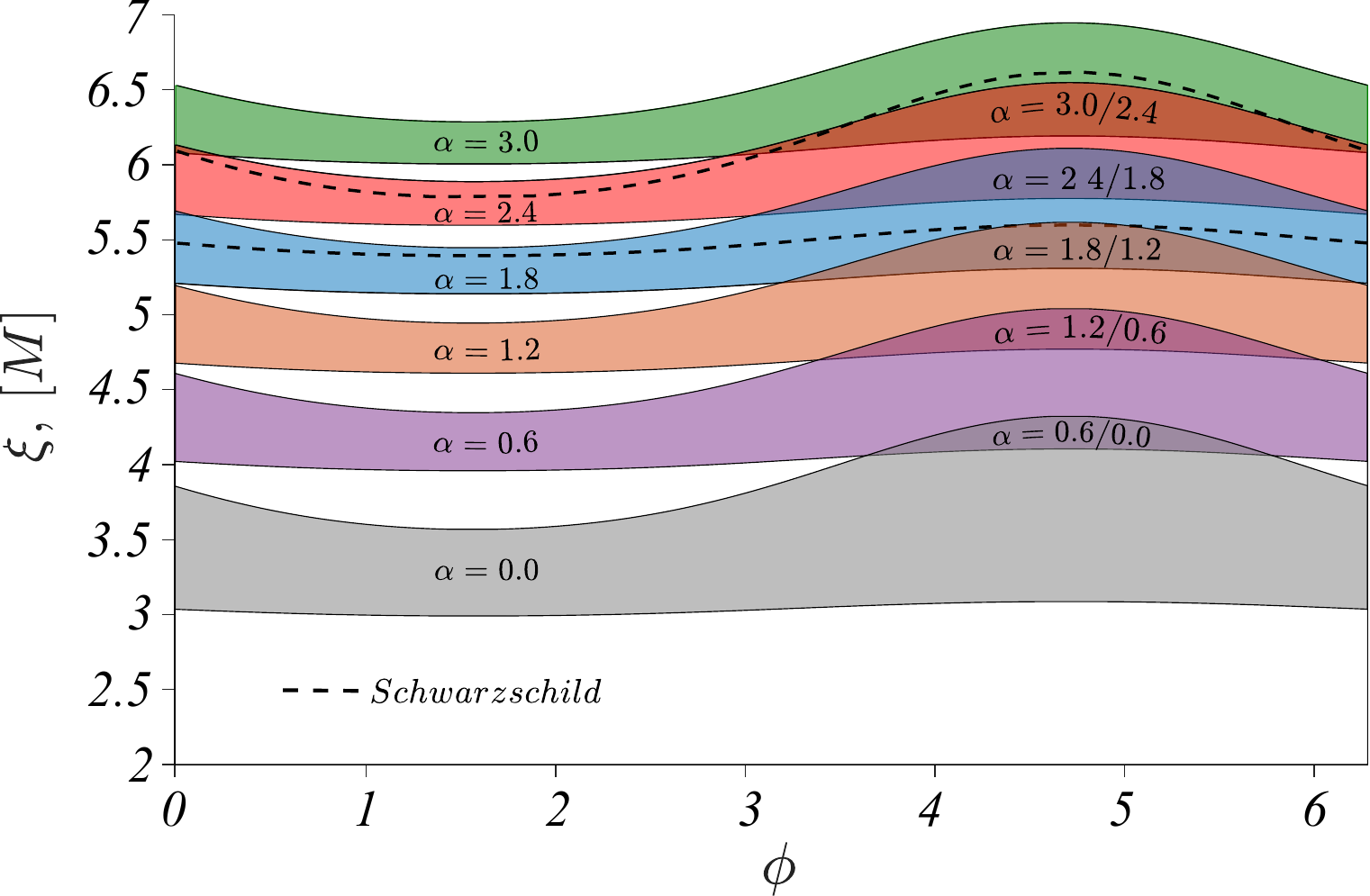}
 \caption{Range of the impact parameter for the indirect image of the thin disk for wormholes with different redshift parameter $\alpha$ and the Schwarzschild black hole (black dashed line). The lower boundary represents the image of the ISCO, while the upper limit corresponds to the circular orbit at $r=50M$.  The inclination angle is $\theta = 20^\circ$.}
\label{fig:disk_k1}
\end{figure}

Due to this property  there exists no overlap between the indirect images of the thin disk around the Schwarzschild black hole and the wormholes with certain values of the redshift parameter $\alpha$. In Fig. $\ref{fig:disk_k1}$ we present the extend of the thin disk image of order  $k=1$ in the impact parameter space for wormholes with various value of $\alpha$ and for the Schwarzschild black hole at the inclination angle  $\theta = 20^\circ$. In particular, in each case we plot the range of the impact parameter $\xi$ which represents the effective radius of each point of the image on the observer's sky\footnote{For an asymptotic observer the impact parameter $\xi$ is expressed by means of the celestial coordinates as $\xi = \sqrt{x^2 + y^2}$.}. The lower boundary for each $\alpha$ represents the image of the ISCO for the corresponding solution, while the upper boundary corresponds to the orbit at $r=50M$. Orbits with larger radii lead to images infinitesimally close to the upper boundary.

\begin{figure}[h!]
\centering
   \includegraphics[width=0.6\textwidth]{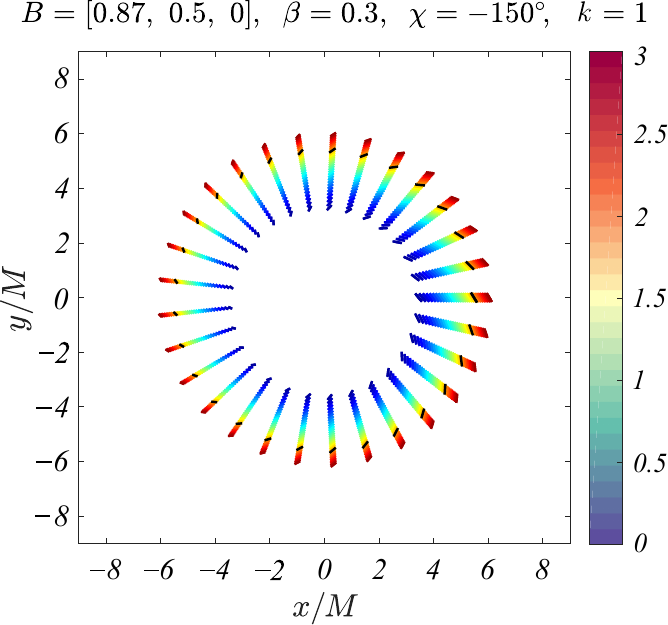}
 \caption{Polarization of the indirect images of order $k=1$ for wormholes with different redshift parameter $\alpha$. Each color represents the observable polarization of the orbit located at $r=6M$ for a particular wormhole solution with  $\alpha\in[0,3]$. The polarization for the Schwarzschild black hole is given by a black dotted line as a reference. The inclination angle is $\theta = 20^\circ$.}
\label{fig:pol_k1}
\end{figure}

We see that the indirect images of the thin disk around wormholes intersect with the thin disk image around the Schwarzschild black hole only for a very narrow range of the redshift parameter close to the value $\alpha = 1.8$. Thus, outside this range we cannot compare the polarization which is observed at a given point of the celestial sphere as a result of the gravitational lensing in a wormhole or black hole spacetime. However, these wormhole solutions can be distinguished by the different apparent radius of the indirect disk image with respect to the Schwarzschild black hole.

The observable polarization for such wormholes is presented in Fig. $\ref{fig:pol_k1}$ for a continuous distribution of the redshift parameter. For each value of $\alpha \in [0,3]$ we illustrate the polarized image of the circular orbit located at $r=6M$. We see that the polarization pattern resembles the Schwarzschild black hole. However, the observable radius of the orbit can deviate considerably.

\begin{figure}[t]
   \centering
   \includegraphics[width=0.99\textwidth]{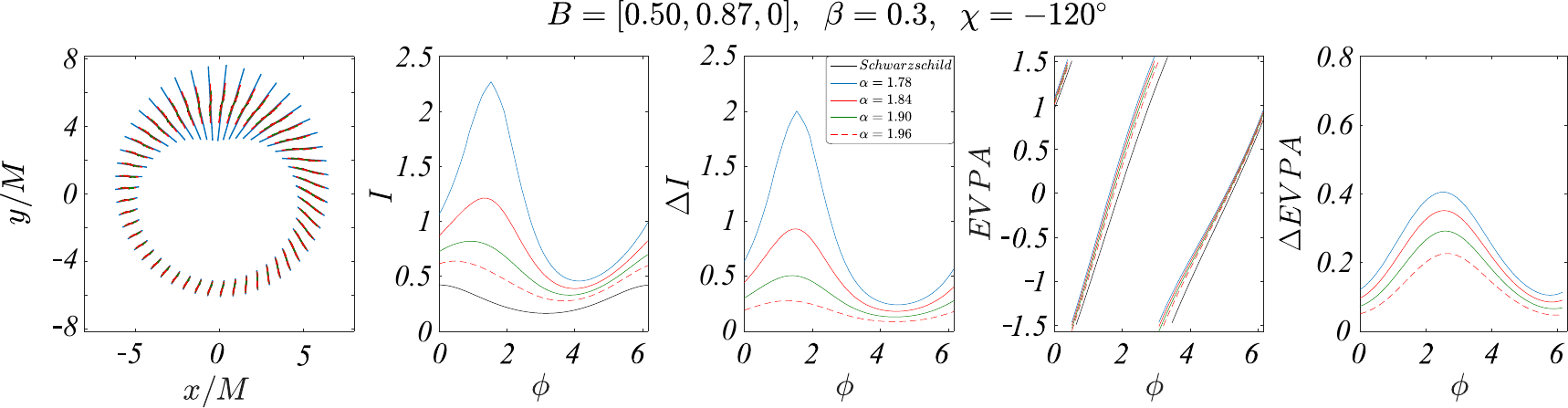}
 \\[2mm]
  \includegraphics[width=0.99\textwidth]{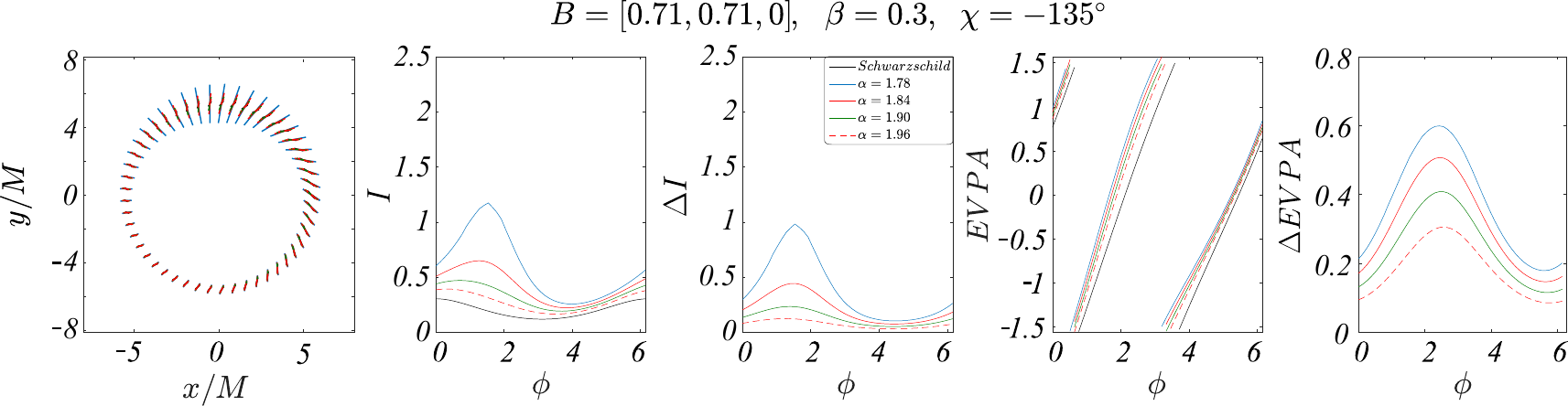} \\[2mm]
  \includegraphics[width=0.99\textwidth]{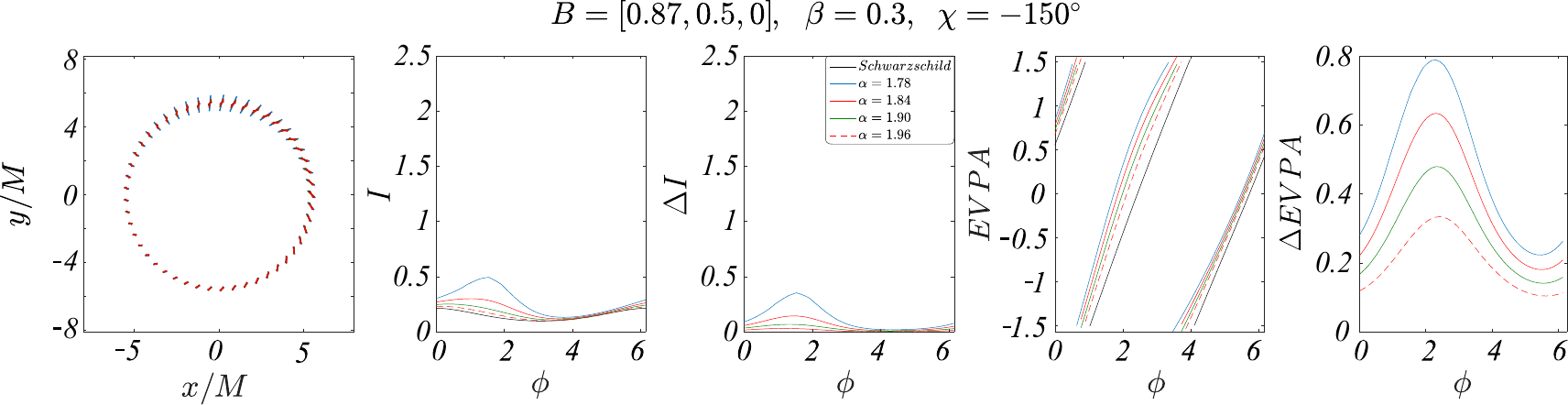}
  \caption{ Polarization of the indirect images of order $k=1$ for wormholes at the inclination angle $\theta = 20^\circ$. We analyze the polarization intensity I and direction EVPA as a function of the redshift parameter $\alpha$, as well as their deviation from the Schwarzschild black hole $\Delta$I and $\Delta$EVPA (see main text).}
\label{fig:pol1_k1}
\end{figure}

In order to access quantitative effects, in Fig. $\ref{fig:pol1_k1}$ we study the observable polarization for wormholes which possess certain overlap of their indirect disk images with the Schwarzschild solution. Thus, we can compare the polarization properties at a given point of the celestial sphere for the two types of solutions.  As in our previous analysis, we present the polarization which is observed at the location of the lensed image of the ISCO for the Schwarzschild black hole. This is possile for a limited range of the redshift parameter $\alpha \in [1.78, 1.96]$. We see that the strong gravitational lensing leads to much more pronounced deviation from the polarization properties of the Schwarzschild black hole. The polarization intensity can deviate considerably getting more than $8$ times larger for the magnetic field direction $B = [0.5, 0.87, 0]$ and $\alpha = 1.78$.  The maximum deviation in the intensity in the image decreases when the radial component of the magnetic field increases. For example, for the magnetic field $B = [0.87, 0.5, 0]$ and $\alpha = 1.78$ it becomes approximately $3.5$ times larger than for the Schwarzschild black hole. The deviation in the polarization direction $\Delta$EVPA possesses the contrary behavior. It is largest for predominantly radial magnetic fields and declines when the azimuthal component of the magnetic field increases. The maximum deviation in the polarization intensity and direction $\Delta$EVPA is presented in Table $\ref{table:theta20_1}$ for the two boundary value of $\alpha$.

\begin{table}[t]
    \centering
      \begin{tabular}{||c|c|c|c||}
       \hline
         \thead{ Magnetic field }   & \thead{$\frac{\text{max}\,\Delta \text{I}}{\text{I}_\text{Sch}} $\,\,\,}
          &\thead{$\frac{\text{max}\,\Delta \text{EVPA}}{\text{EVPA}_\text{Sch}}$\,\,\,}&   $\alpha$   \\    \hline
           \thead{$\text{B = [0.5, 0.87, 0]}$}  &  \thead{7.5 \\[2mm] 0.93} & \thead{0.53 \\[2mm] 0.25} &\thead{1.78\\[2mm]1.96}
          \\  \hline
          \thead{$\text{B= [0.71, 0.71, 0]}$ } &  \thead{5.1 \\[2mm] 0.58} & \thead{1.8 \\[2mm] 0.64} &  \thead{1.78\\[2mm]1.96}
          \\  \hline
           \thead{$\text{B= [0.87, 0.5, 0]}$}  &  \thead{2.5 \\[2mm] 0.18} & \thead{6.5 \\[2mm] 38 } &  \thead{1.78\\[2mm]1.96}
          \\  \hline
       \end{tabular}
      \caption{Deviation of the wormhole polarization from the Schwarzschild black hole for the  indirect images of order $k=1$ and inclination angle $\theta = 20^\circ$. In each case we give  the maximum relative deviations $\text{max}\,\Delta\text{I}/\text{I}_\text{Sch}$ and $\text{max}\,\Delta\text{EVPA}/\text{EVPA}_\text{Sch}$   with respect to the  Schwarzschild solution, which are reached in the polarized image.}
    \label{table:theta20_1}
\end{table}

We further observe that the intensity around the ring changes its distribution with respect to the Schwarzschild black hole. When $\alpha$ decreases the maximum of the observable intensity moves upwards at the right-hand side of the image. Thus, for $\alpha = 1.96$ it is located approximately at the same azimuthal angle as for the Schwarzschild solution while for $\alpha = 1.78$ it already reaches the top of the image. This qualitative effect combined with the considerable deviation in the intensity magnitude at predominantly azimuthal magnetic field can serve as an observational signature for distinguishing wormhole spacetimes from the Schwarzschild black hole.


\subsection{Double-disk wormholes}

In this section we consider the possibility that the wormholes possess accretion disks on both sides of their throats and signals from the region beyond the throat are able to reach our universe in a reasonable time and get detected by an asymptotic observer. We can further assume the matter accretes in similar physical conditions  on both sides of the throat and emits linearly polarized radiation. In this case, the synchrotron radiation from the region beyond the throat will have impact on the polarized image of the disk on the observer's sky in our universe.

\begin{figure}[t]
\centering
\begin{tabular}{cc}
   \includegraphics[width=0.99\textwidth]{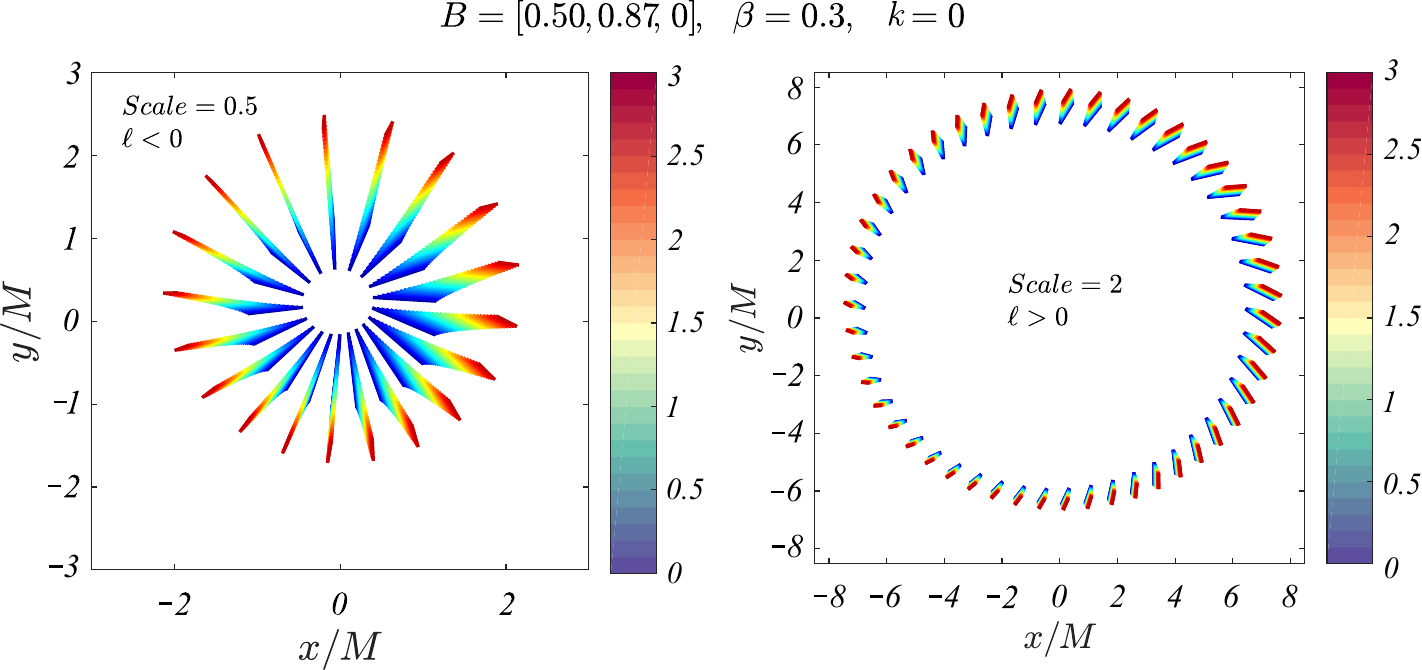} \\
    $a)$  \hspace{6.5cm} $b)$
   \end{tabular}
 \caption{Polarized images for wormholes with accretion disks on both sides of the throat. In b) we represent the observable polarization for the orbit at $r=6M$ in our universe, while a) is the image of the orbit located symmetrically at the same radial distance on the other side of the throat. Each color represents a particular wormhole solution with  $\alpha\in[0,3]$. The polarization intensity is represented with a different scale factor in the two images. Thus, a line element in a) is in fact 4 times larger than a line element in b). The inclination angle is $\theta = 20^\circ$.}
\label{fig:pol_double}
\end{figure}

We explore the contribution of the second accretion disk by considering the polarized image of a magnetized fluid ring located in the region beyond the wormhole throat. For the purpose we extend the wormhole metric ($\ref{metric}$) across the throat by introducing a global radial coordinate $l$ defined as

\begin{eqnarray}
dl = \pm \frac{dr}{\sqrt{1-\frac{r_0}{r}}},
\end{eqnarray}or explicitly

\begin{eqnarray}
l = \pm\left[\sqrt{\frac{r}{r_0}\left(\frac{r}{r_0}-1\right)} + \ln\left( \sqrt{\frac{r}{r_0}} + \sqrt{\frac{r}{r_0}-1}  \right)\right].
\end{eqnarray}

The coordinate $l$ takes the range $-\infty<l< +\infty$ as
the limits $r\rightarrow -\infty$ and $r\rightarrow + \infty$ correspond to the two asymptotic ends, while the wormhole throat is located at $l=0$. In this way the metric becomes regular at the wormhole throat

\begin{equation}
    ds^2 = -N^2(l)dt^2 +dl^2 +r^2(l)\left(d\theta^2+ \sin^2\theta d\phi^2\right),
   \end{equation}
and describes smoothly the transition between the two asymptotically flat regions for positive and negative values of $l$.

In Fig. $\ref{fig:pol_double}$ we present the polarized image of the circular orbit at $r=6M$ in our universe and the image of the orbit located symmetrically at the same radial distance with respect to the throat, however in the other asymptotically flat region. We see that the radiation which originates from the region beyond the throat possesses rather distinct polarization properties. The polarization pattern changes since we observe a less pronounced twist of the polarization direction around the ring. The polarization is almost radial in the left-hand side of the image and slightly deviates from the radial direction at the right-hand side. The deviation from the radial distribution depends on the redshift parameter as it decreases for small $\alpha$. The polarization intensity increases substantially compared to  the radiation from the asymptotically flat region for positive $l$. Depending on the value of $\alpha$ it grows approximately from 3.5 to 6.5 times in magnitude. Thus, the radiation from the asymptotically flat region across the throat will contribute substantially to the observed polarized image and provide a characteristic signature for the detection of wormhole geometry. We will discuss these effects in more detail for different magnetic field configurations in a further work.

\section{Conclusion}

In this work we studied the linear polarization of the emission from the accretion disk around a class of static traversable wormholes. Our aim is to investigate the influence of the spacetime metric on the observable polarized images and explore the possibility for detecting exotic compact objects with polarization experiments. Using the simplified model of a magnetized fluid ring which orbits in the equatorial plane and emits synchrotron radiation, we simulate the observable polarization in wormhole geometries for a range of physical parameters and compare with the Schwarzschild black hole. In order to be able to reproduce the observed polarization of M87*, we focus mainly on equatorial magnetic fields in our analysis.

We consider three types of polarized images with different properties. Initially, we study the weakly lensed direct images of orbits in the region of the strong gravitational interaction for small and large inclination angles. In this case the  polarization pattern is qualitatively very similar  for wormhole and black hole spacetimes. For our sample of physical parameters the deviation in the polarization intensity between the two types of compact objects is less than $43\%$ for small inclination angles, while  the polarization direction EVPA deviates less than $3.8\%$. Moreover, by fine-tuning the wormhole geometry it is possible to obtain horizonless spacetimes which closely mimic the Schwarzschild black hole in its polarization properties.  Thus, we can construct wormholes which deviate from the Schwarzschild solution  less than $4\%$ in their polarization intensity and  less than $3\%$ in  EVPA for a range of equatorial magnetic fields. Increasing the inclination angle the distinction between the two types of compact objects grows. For observer located at the inclination angle $\theta =70^\circ$ we estimate that the polarization intensity can increase approximately 2 times with respect to the Schwarzschild solution,  while the deviation of the polarization direction is within $25\%$.

Next, we consider the strongly lensed indirect images which lead to more pronounced phenomenological effects. We demonstrate that for images of the order $k=1$ the deviation in the polarization properties for small inclination angles can grow with an order in magnitude. The polarization intensity in wormhole spacetimes can become more than eight times larger than for the Schwarzschild black hole, while the deviation in EVPA can reach $50\%$. In addition, the maximum of the polarization intensity shifts upwards towards the top of the image compared to the Schwarzschild black hole. This qualitative effect combined with the considerable deviation in the intensity magnitude  could serve as an observational signature for distinguishing wormhole spacetimes.

Finally, we assume that the wormhole possesses accretion disks on both sides of its throat and we can detect  radiation from the asymptotically flat region across the throat. We see that the polarized radiation which reaches the observer through the throat can become more than six times more intensive compared to the radiation in our universe and it forms an additional structure of ring-like images at smaller observational radii. The twist of the polarization direction around these rings is less pronounced and the polarization pattern changes. Thus,  the  properties of the polarized images of the radiation across the throat are significantly different, and  can be considered as a characteristic signature for detecting horizonless objects.

Based on the analysis of our class of wormhole geometries we conclude that at small inclination angles it could be difficult to distinguish wormhole from black hole spacetimes by their direct polarized images. Strongly lensed indirect images provide more reliable probes of the underlying spacetime, as well as characteristic effects such as the detection of the polarized radiation from the region across the wormhole throat.



\section{Appendix}

\subsection*{A. Lensing properties of the circular orbits}

\begin{figure}[h]
   \centering
   \includegraphics[width=0.99\textwidth]{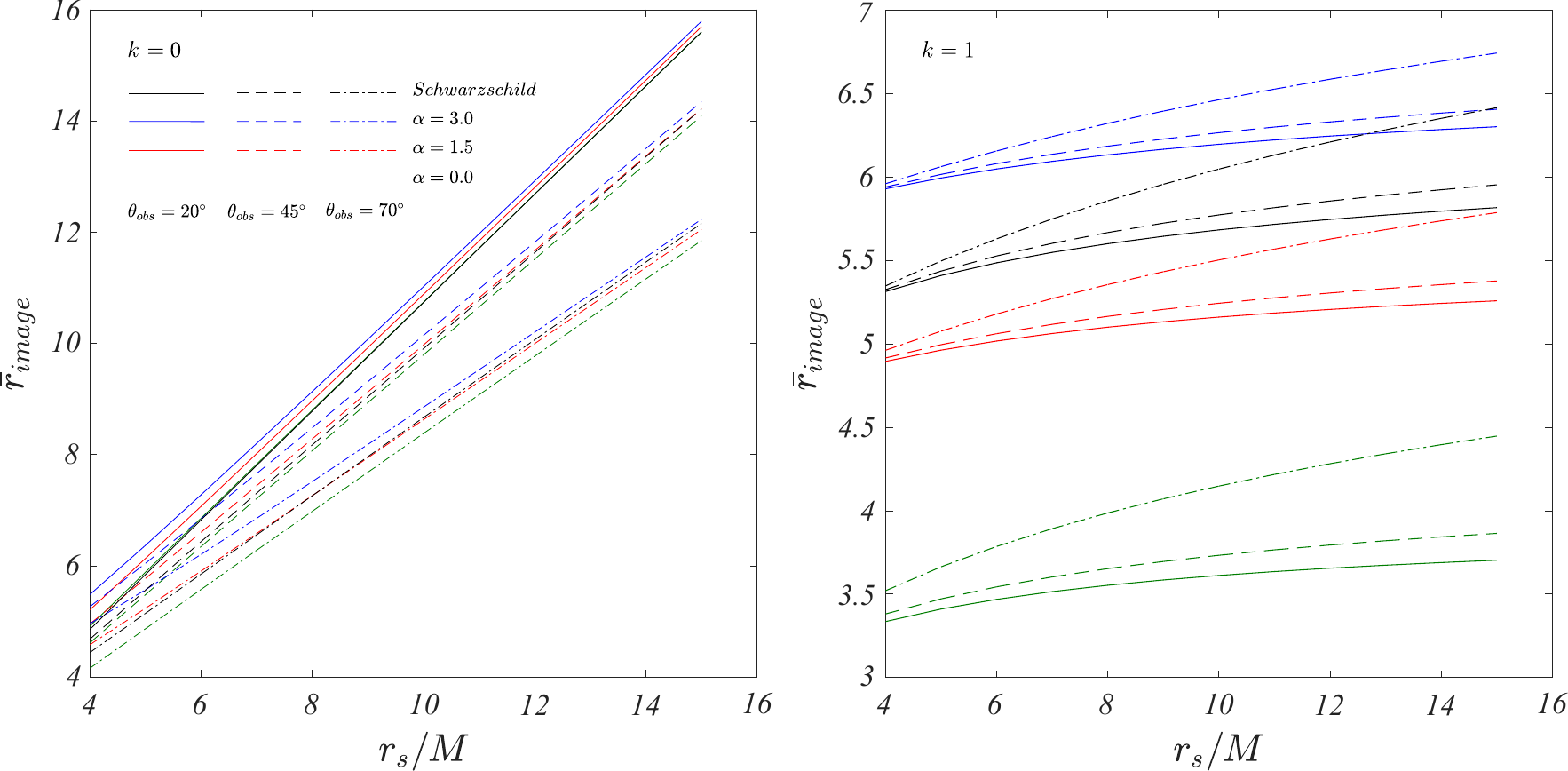}
   \caption{Effective radius of the lensed image of the equatorial circular orbits for direct (left) and indirect (right) images.}
   \label{fig:lens_map}
\end{figure}

We investigate the apparent radius of the  equatorial circular orbits at the observer's sky as a function of the wormhole redshift parameter $\alpha$ and the inclination angle $\theta$ in comparison to the Schwarzschild black hole.  Since the radial position of each point of the lensed image depends on the azimuthal angle we define the effective radius
\begin{eqnarray}\label{r_image}
\bar r_{image} = \frac{1}{4}[(x_{max} - x_{min})|_{y=0} + (y_{max} - y_{min})|_{x=0} ],
\end{eqnarray}
where we consider the maximal deviations in the image along the $x$- and $y$-axes. For small inclination angles the images are close to circular and eq. ($\ref{r_image}$) gives a reasonable approximation for their radii. For larger inclination angles the images are considerably squashed in the vertical direction. Still, we can consider the quantity $r_{image}$ as a measure for comparing the lensing properties of the wormhole solutions and the Schwarzschild black hole. In Fig. $\ref{fig:lens_map}$ we see that the circular orbits for the two types of compact objects are visualized at similar radial locations at the observer's sky for all the inclination angles and redshift parameters. We should note however that we consider here only the direct images. The strong gravitational lensing causes the apparent location of the indirect images to depend substantially on the redshift parameter (see Fig. $\ref{fig:disk_k1}$ in section 4.2).

\subsection*{B. Direct polarized images: examples}

\begin{figure}[htbp]
   \centering
   \includegraphics[width=0.99\textwidth]{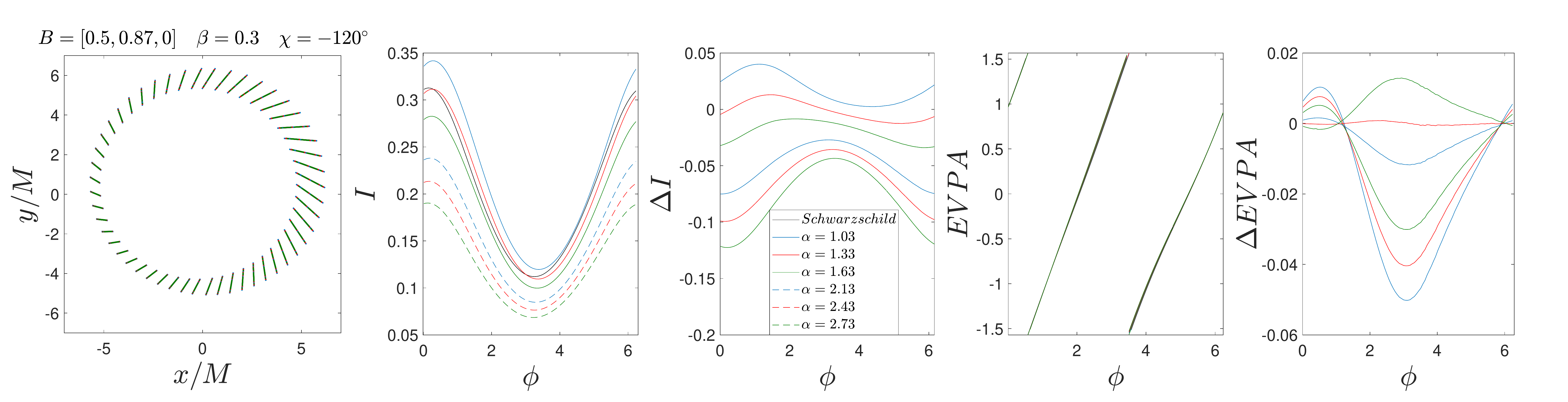}\\[1mm]
   \includegraphics[width=0.99\textwidth]{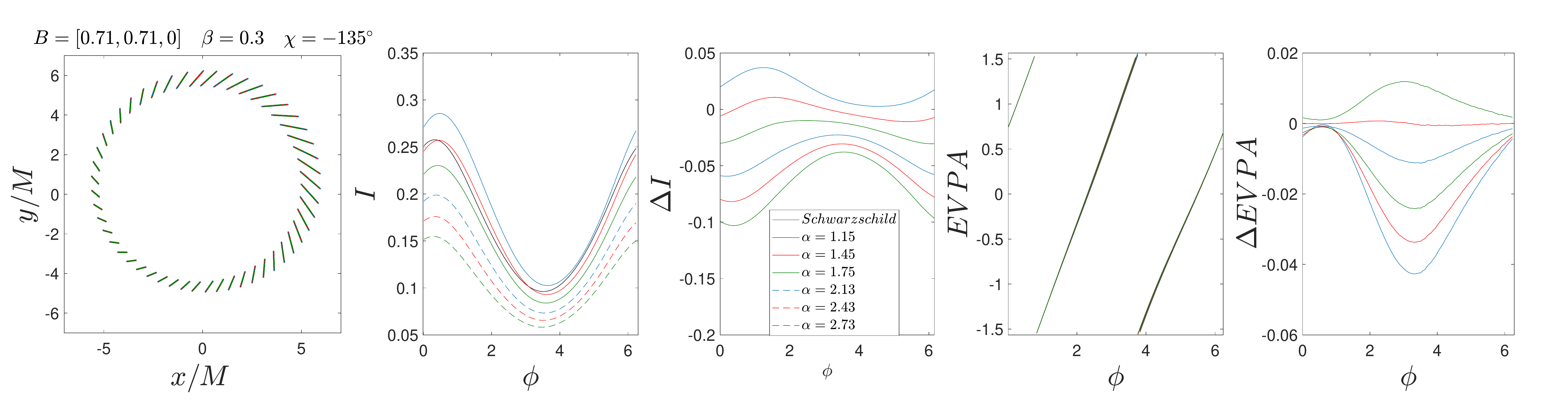}\\[1mm]
   \includegraphics[width=0.99\textwidth]{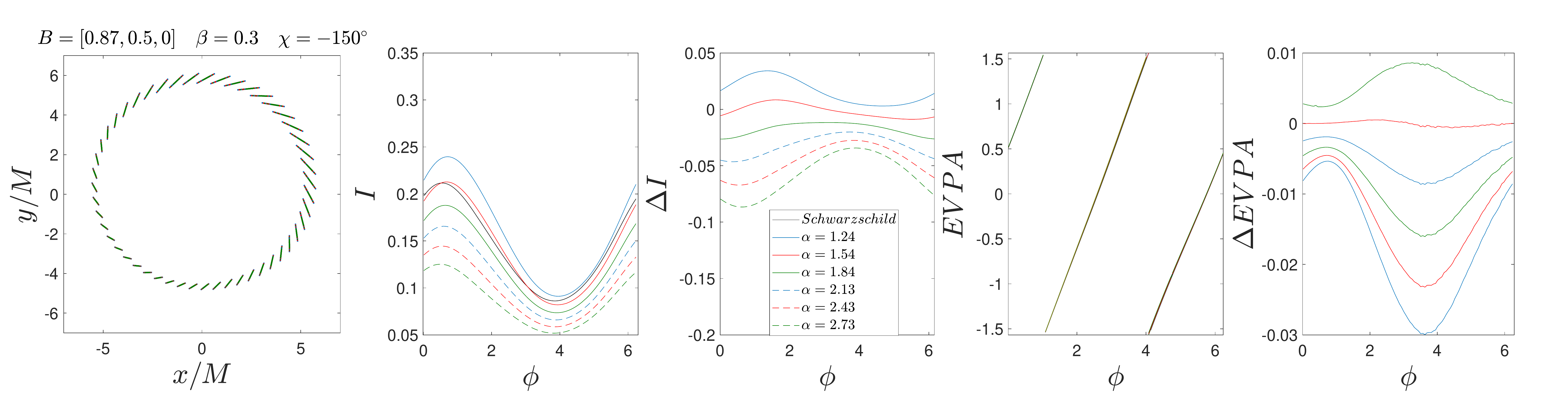}
\caption{Linear polarization for static wormholes at the inclination angle $\theta =20^\circ$. We analyze the polarization intensity I and direction EVPA as a function of the redshift parameter $\alpha$, as well as their deviation from the Schwarzschild black hole $\Delta$I and  $\Delta$EVPA. The two critical value of $\alpha$, which lead to minimal deviation $\Delta$I and  $\Delta$EVPA  are presented in red solid and red dashed lines, respectively.}
\label{fig:polarization_20_1}
\end{figure}

\begin{figure}[]
\centering
 \subfloat[][]{
  \includegraphics[width=0.75\textwidth]{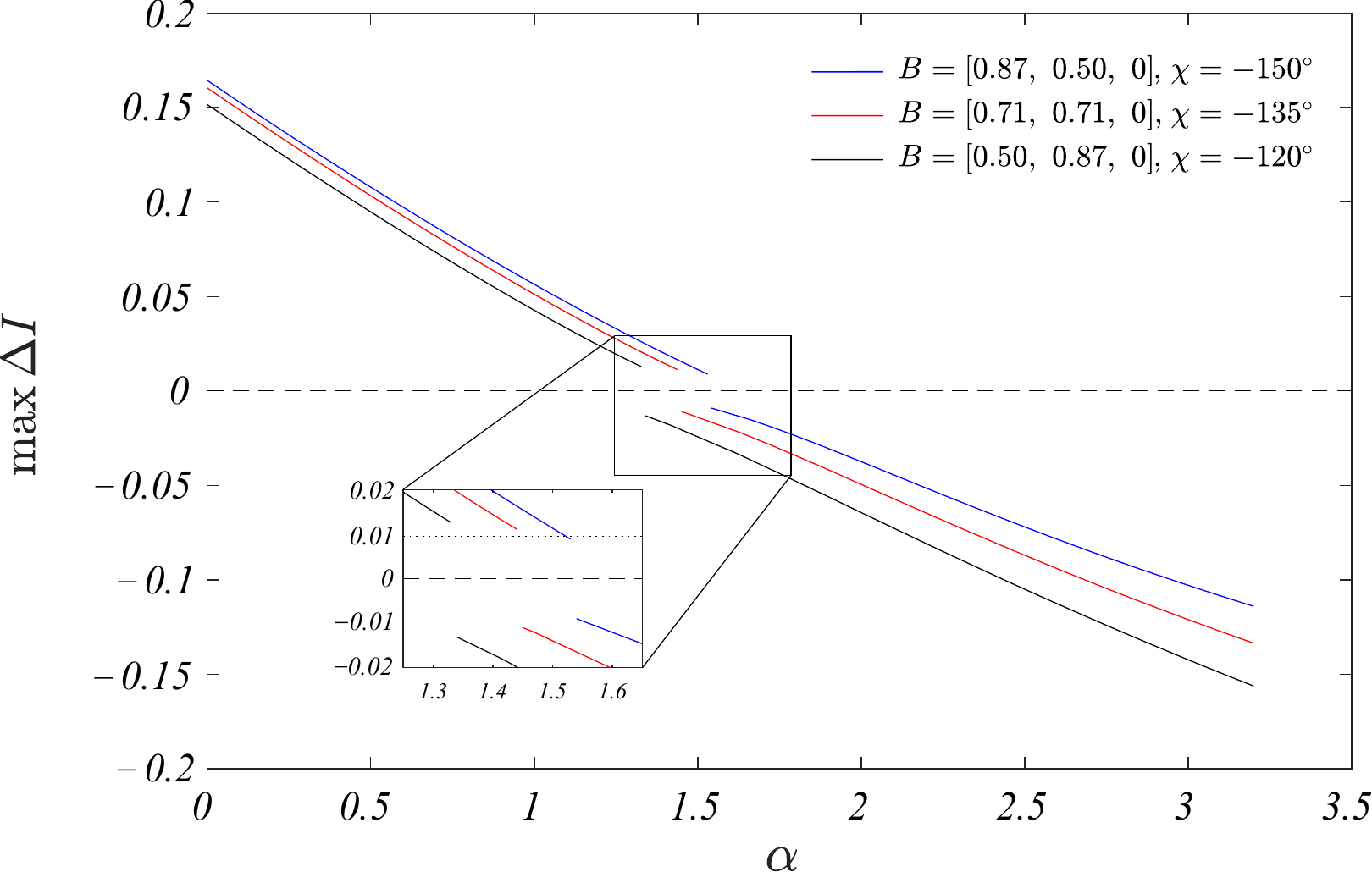}
 } \\[12mm]
  \subfloat[][]{
  \includegraphics[width=0.75\textwidth]{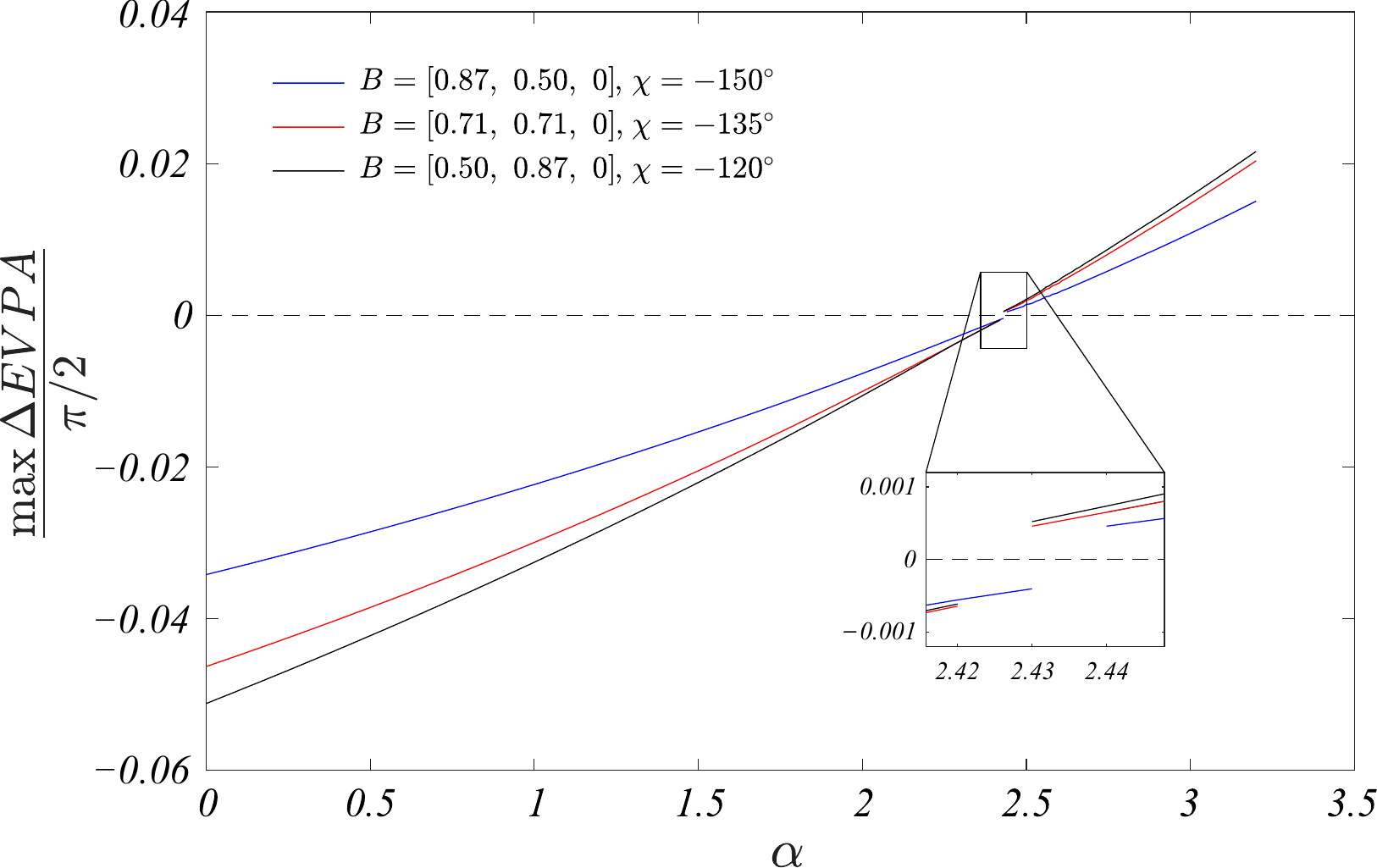}
 }
\caption{Maximum deviation of the polarization properties of the wormhole from the Schwarzschild black hole as a function of the redshift parameter $\alpha$ for the inclination angle $\theta = 20^\circ$. We plot the maximum deviation of the intensity max $\Delta$I (a), and the polarization direction max $\Delta$EVPA (b), which is reached in the images presented in Fig. $\ref{fig:polarization_20_1}$  for each value of $\alpha$. Negative values imply that the corresponding quantity is larger for the Schwarzschild solution than for the wormhole.}
\label{fig:maxI_20_1}
\end{figure}

\begin{figure}[]
\centering
   \includegraphics[width=0.99\textwidth]{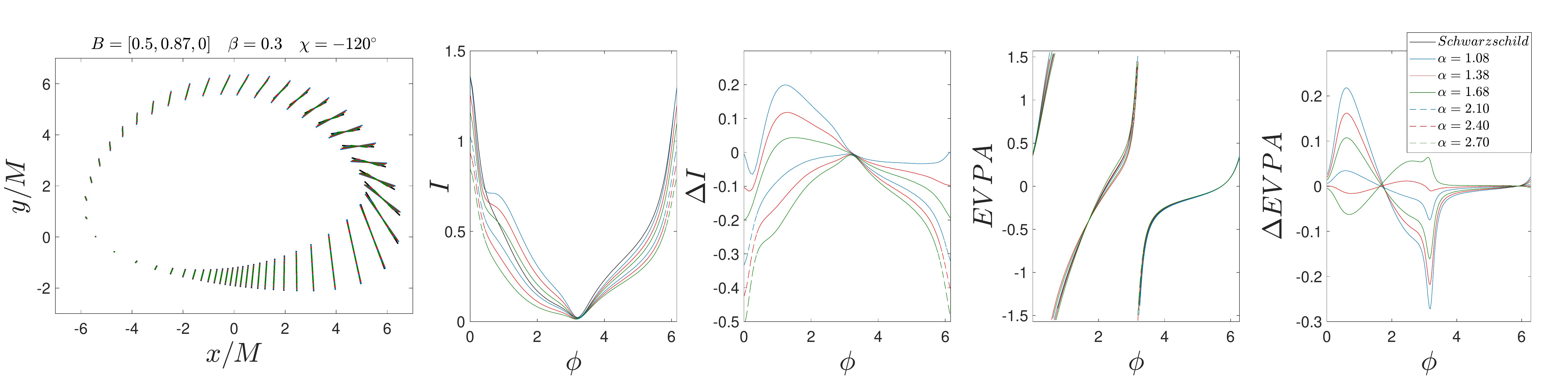}\\[2mm]
  \includegraphics[width=0.99\textwidth]{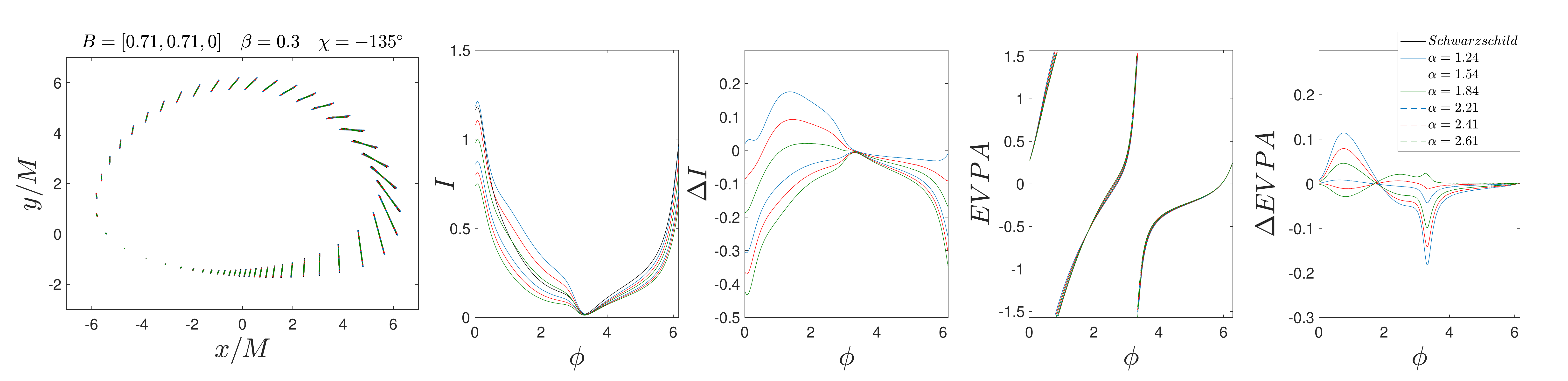} \\[2mm]
  \includegraphics[width=0.99\textwidth]{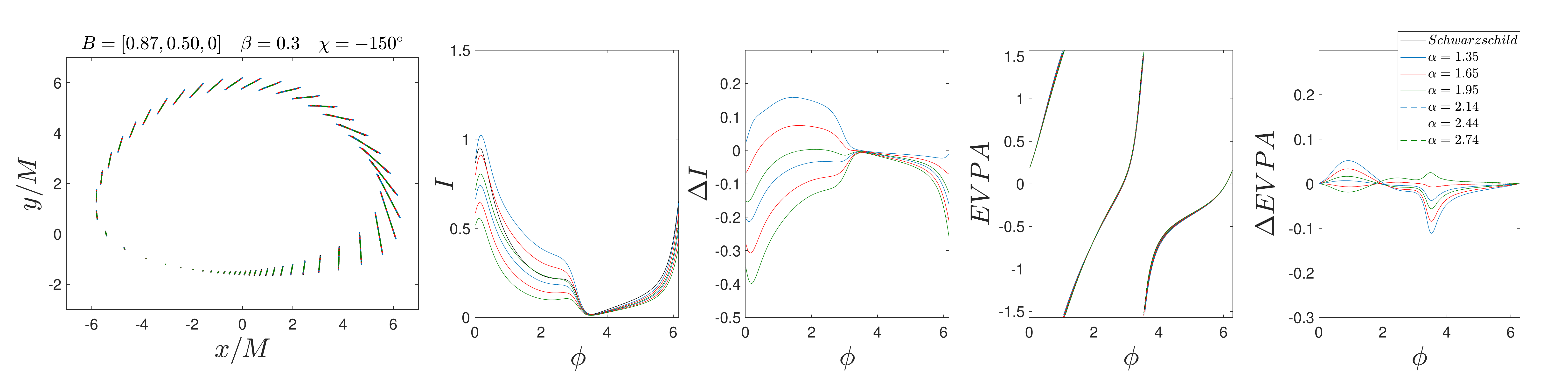}
 \caption{Linear polarization for static wormholes at the inclination angle $\theta =70^\circ$. We follow the same conventions as in Fig. $\ref{fig:polarization_20_1}$. }
\label{fig:polarization_70_1}
\end{figure}

\begin{figure}[]
\centering
 \subfloat[][]{
  \includegraphics[width=0.75\textwidth]{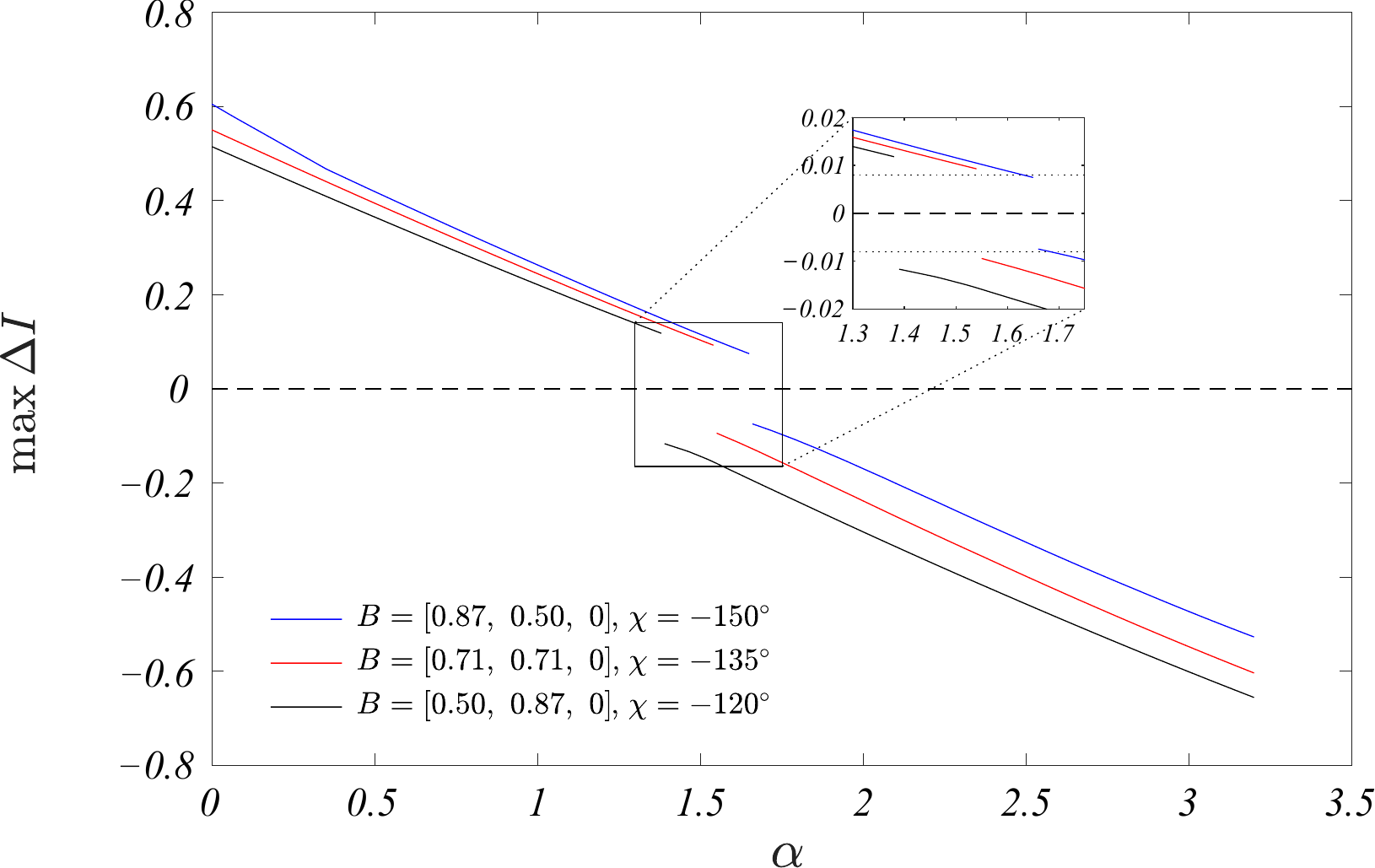}
 } \\[12mm]
  \subfloat[][]{
  \includegraphics[width=0.75\textwidth]{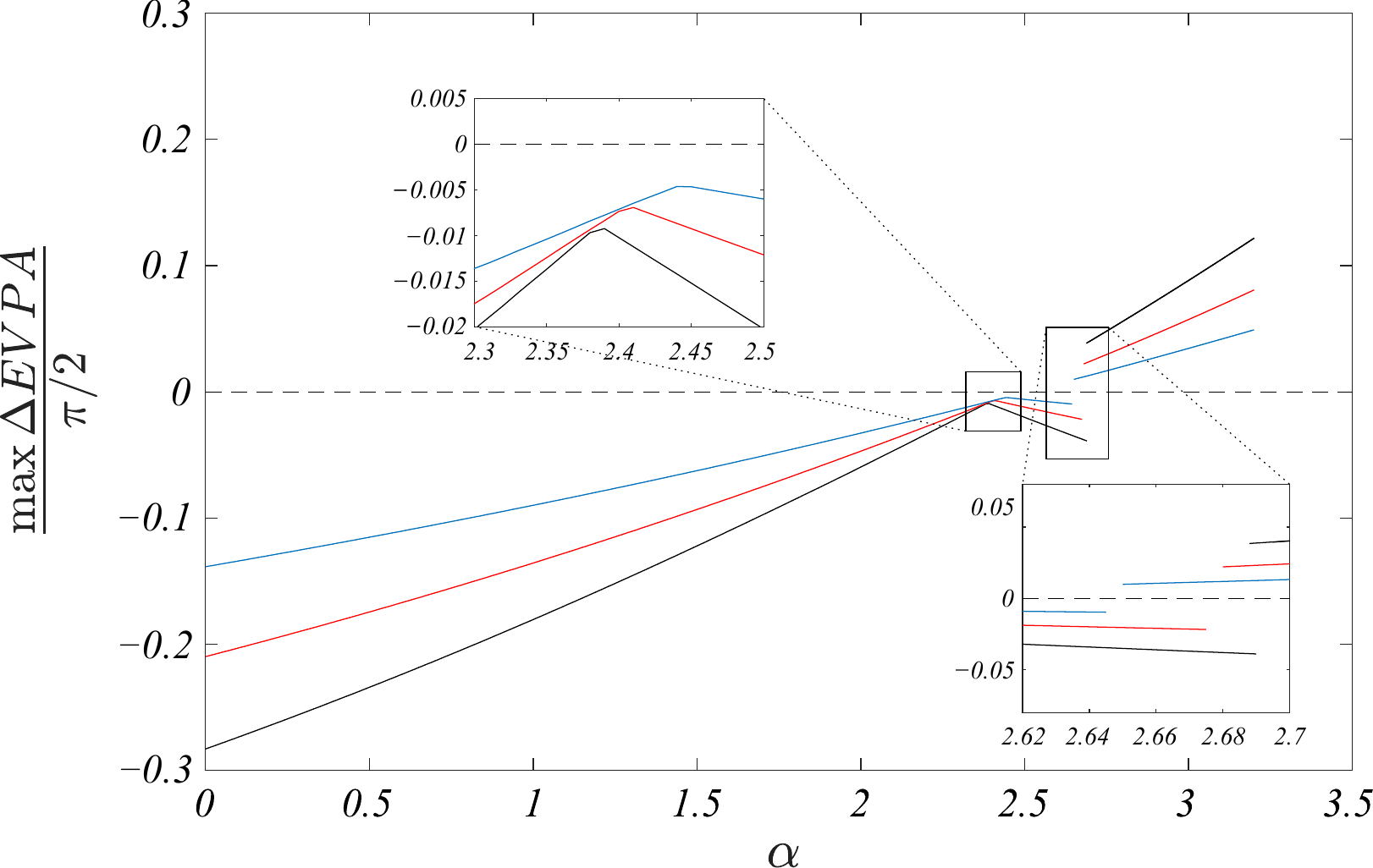}
 }
\caption{Maximum deviation of the polarization properties of the wormhole from the Schwarzschild black hole as a function of the redshift parameter $\alpha$ for the inclination angle $\theta = 70^\circ$. We follow the same conventions as in Fig. \ref{fig:maxI_20_1}.}
\label{fig:maxI_70_1}
\end{figure}

Here we give further examples of the observable polarization for wormholes in comparison to the Schwarzschild black hole. In particular we explore the influence of the emission radius on the polarization pattern by considering magnetized fluid rings which are located deeper in the gravitational field of the compact objects. In Figs. $\ref{fig:polarization_20_1}$ - $\ref{fig:maxI_70_1}$ we visualize the wormhole polarization at the observable location of the orbit with radius $r=4.5M$ for the Schwarzschild black hole and analyse its properties following the discussion in section 4.1. We consider purely equatorial magnetic fields and two cases of different inclination angles.

\section*{Acknowledgments}
We gratefully acknowledge support by the Bulgarian NSF Grant KP-06-H38/2.

\end{document}